\documentclass[acmsmall]{acmart}
\usepackage{subcaption} %% For complex figures with subfigures/subcaptions
\usepackage{booktabs}   %% For formal tables:
\usepackage{subcaption} %% For complex figures with subfigures/subcaptions
\usepackage{qcircuit}
\usepackage{rotating} 
\usepackage{url}
\usepackage{proof}
\usepackage{bbm}
\usepackage{macros}
\usepackage{lstcoq}
\usepackage{mathtools}
\usepackage{xcolor}
\usepackage{amsmath}
\usepackage{physics}
\usepackage{multicol}
\usepackage{enumitem}
\usepackage[ruled,vlined]{algorithm2e}
\usepackage{mathpartir}
\usepackage{amsthm,mathrsfs}
\usepackage{amsmath,amsthm,bbold, dsfont}
\usepackage{mathtools}
\usepackage{pgf, tikz}
\usepackage{pgfplots}
\usetikzlibrary{arrows, automata}
\usepackage{graphicx}
\usepackage{color}
\usepackage{tikz-cd}
\usepackage{geometry}
\usepackage{multirow} 
\usepackage{array}
\usepackage{listings}
\definecolor{codegreen}{rgb}{0,0.6,0}
\definecolor{codegray}{rgb}{0.5,0.5,0.5}
\definecolor{codepurple}{rgb}{0.58,0,0.82}
\definecolor{backcolour}{rgb}{0.95,0.95,0.92}
\lstdefinestyle{mystyle}{
    keywordstyle=\color{magenta},
        breaklines=true, 
        backgroundcolor=\color[RGB]{245,245,244},
    captionpos=b,   
    basicstyle=\scriptsize\ttfamily,
    numberstyle=\tiny\color{gray},
    keepspaces=true,                 
    numbersep=5pt,                  
    showspaces=false,       
    showstringspaces=false,
    showtabs=false,                  
    tabsize=2,
    escapeinside={@}{@},
}
\newcommand{\keynum}[1]{\textcolor{blue}{#1}}

\newtheorem{definition}{Definition}[section]
\newtheorem{theorem}{Theorem}[section]
\newtheorem{example}{Example}[section]

\theoremstyle{definition}
\theoremstyle{remark}

\usepackage{makecell}
\newcolumntype{C}[1]{>{\centering\arraybackslash}p{#1}}

\newcommand{\X}{\mathcal{X}}
\newcommand{\Y}{\mathcal{Y}}

\DeclareGraphicsExtensions{.jpg,.pdf}

\newcommand{\opn}[1]{\operatorname{#1}}

\definecolor{webgreen}{rgb}{0,.5,0}
\definecolor{webblue}{rgb}{0,0,.5}

\def\o{\opn{o}}

\usepackage{proof}

\usepackage[matrix,frame,arrow]{xypic}
\usepackage{qcircuit}

\renewcommand{\ket}[1]{|#1\rangle} %
\renewcommand{\bra}[1]{\langle #1|} %
\usepackage{lstcoq}
\usepackage{proof}
\usepackage{bbm}
\usepackage{macros}
\usepackage{lstcoq}
\usepackage{mathtools}
\usepackage{xcolor}
\usepackage{amsmath}

\usepackage{multicol}
\usepackage{enumitem}
\usepackage[ruled,vlined]{algorithm2e}
\usepackage{wrapfig}
\usepackage{amsthm,mathrsfs}
\newcommand\utimes{\mathbin{\ooalign{$\cup$\cr%
   \hfil\raise0.42ex\hbox{$\scriptscriptstyle\times$}\hfil\cr}}}

\newcommand{\interpret}[3]{#1\triangleright #2\leftrightarrow#3}

\newcommand{\ufeq}[3]{#2 \equiv^{#1} #3}

\newcommand{\phase}[1]{e^{2\pi i\cdot d(#1)}}
\newcommand{\dvexp}[2]{#1\cdot \phase{#2}}
\newcommand{\bvalue}[1]{\delta(#1)}

\newcommand{\const}[1]{\textbf{const}~\text{#1}}
\newcommand{\relabel}[2]{\textbf{relabel}~#1~#2}

\newcommand{\fix}{\textbf{fix}}
\newcommand{\seq}[2]{\textbf{seq}~#1~#2}
\newcommand{\shift}[3]{\textbf{shift}~#1~#2~#3}
\newcommand{\permute}[2]{w[#1 \rightleftharpoons #2]}
\newcommand{\mask}{\text{\textbf{mk}}}
\newcommand{\sparse}[3]{#1\unlhd (#2,#3)}

\usepackage{xspace}
\usepackage{enumitem}
\newcommand{\sqir}[0]{\textsc{SQIR}\xspace}
\newcommand{\sqirs}[0]{\textsc{ISQIR}\xspace}
\newcommand{\ssem}[1]{\{\!\!\{ #1 \}\!\!\}}
\newcommand{\sem}[1]{[\!\![ #1 ]\!\!]}

\newcommand{\app}[1]{\hyperref[app:#1]{Appendix~\ref*{app:#1}}}
\newcommand{\theom}[1]{\hyperref[thm:#1]{Theorem~\ref*{thm:#1}}}

\newcommand{\fig}[1]{\hyperref[fig:#1]{Figure~\ref*{fig:#1}}}

\newcommand{\name}{QSynth\xspace}
\newcommand{\logicName}{hypothesis-amplitude\xspace}
\newcommand{\nat}{\mathbb{N}}
\newcommand{\Cref}[1]{Fig~\ref{#1}}
\newcommand{\amplfunc}{PPSA\xspace}
\renewcommand{\coqe}[1]{\texttt{#1}}

\newcommand{\numbenchmark}{10\xspace}
\setcopyright{rightsretained}
\acmDOI{10.1145/3632901}
\acmYear{2024}
\copyrightyear{2024}
\acmSubmissionID{popl24main-p282-p}
\acmJournal{PACMPL}
\acmVolume{8}
\acmNumber{POPL}
\acmArticle{59}
\acmMonth{1}
\received{2023-07-11}
\received[accepted]{2023-11-07}

\numberwithin{equation}{section}
\usepackage{bm}
\begin{document}

\title[A Case for Synthesis of Recursive Quantum Unitary Programs]{A Case for Synthesis of Recursive Quantum Unitary Programs}  

\author{Haowei Deng}
\authornote{Both authors contributed equally to the paper.}
\orcid{0000-0003-3616-9337}             %% \orcid is optional
\affiliation{
  %\position{Position1}
  %\department{Department1}              %% \department is recommended
  \institution{University of Maryland, College Park}            %% \institution is required
  %\streetaddress{Street1 Address1}
  \city{College Park}
  \state{MD}
  %\postcode{Post-Code1}
  \country{USA}                    %% \country is recommended
}
\email{hwdeng@umd.edu}   

\author{Runzhou Tao}
\authornotemark[1]
\orcid{0000-0002-3733-5168}             %% \orcid is optional
\affiliation{
  %\position{Position1}
  %\department{Department1}              %% \department is recommended
  \institution{Columbia University}            %% \institution is required
  %\streetaddress{Street1 Address1}
  \city{New York}
  \state{NY}
  %\postcode{Post-Code1}
  \country{USA}                    %% \country is recommended
}
\email{runzhou.tao@columbia.edu}          %% \email is recommended
\author{Yuxiang Peng}
\orcid{0000-0003-0592-7131}             %% \orcid is optional
\affiliation{
  %\position{Position1}
  %\department{Department1}              %% \department is recommended
  \institution{University of Maryland, College Park}            %% \institution is required
  %\streetaddress{Street1 Address1}
  \city{College Park}
  \state{MD}
  %\postcode{Post-Code1}
  \country{USA}                    %% \country is recommended
}
\email{ypeng15@umd.edu} 

\author{Xiaodi Wu}
\orcid{0000-0001-8877-9802}             %% \orcid is optional
\affiliation{
  %\position{Position1}
  %\department{Department1}              %% \department is recommended
  \institution{University of Maryland, College Park}            %% \institution is required
  %\streetaddress{Street1 Address1}
  \city{College Park}
  \state{MD}
  %\postcode{Post-Code1}
  \country{USA}                    %% \country is recommended
}
\email{xwu@cs.umd.edu} 

%% Abstract
\begin{abstract}
Quantum programs are notoriously difficult to code and verify due to unintuitive quantum knowledge associated with quantum programming. Automated tools relieving the tedium and errors associated with low-level quantum details would hence be highly desirable. 
In this paper, we initiate the study of \emph{program synthesis} for quantum unitary programs that recursively define a family of unitary circuits for different input sizes, which are widely used in existing quantum programming languages. 
Specifically, we present \name, the first quantum program synthesis framework,  including a new inductive quantum programming language, its specification, a sound logic for reasoning, and an encoding of the reasoning procedure into SMT instances. 
By leveraging existing SMT solvers, \name successfully synthesizes \numbenchmark quantum unitary programs including quantum arithmetic programs, quantum eigenvalue inversion, quantum teleportation and Quantum Fourier Transformation, which can be readily transpiled to executable programs on major quantum platforms, e.g., Q\#, IBM Qiskit, and AWS Braket. 
%Moreover, we also extend \name to handle quantum programs with oracles, where we succeed in synthesizing quantum programs for the famous Deutsch-Josza problem. 

\end{abstract}
%% 2012 ACM Computing Classification System (CSS) concepts
%% Generate at 'http://dl.acm.org/ccs/ccs.cfm'.

% <concept_id>10011007.10011006.10011008</concept_id>
% <concept_desc>Software and its engineering~General programming languages</concept_desc>
% <concept_id>10003456.10003457.10003521.10003525</concept_id>
% <concept_desc>Social and professional topics~History of programming languages</concept_desc>
% <concept_significance>300</concept_significance>

% \ccsdesc[500]{Software and its engineering~General programming languages}
% \ccsdesc[300]{Social and professional topics~History of programming languages}
% %% End of generated code

%% Keywords
%% comma separated list
\begin{CCSXML}
<ccs2012>
   <concept>
       <concept_id>10011007.10011006.10011050.10011017</concept_id>
       <concept_desc>Software and its engineering~Domain specific languages</concept_desc>
       <concept_significance>500</concept_significance>
       </concept>
   <concept>
       <concept_id>10010583.10010786.10010813.10011726</concept_id>
       <concept_desc>Hardware~Quantum computation</concept_desc>
       <concept_significance>300</concept_significance>
       </concept>
 </ccs2012>
\end{CCSXML}

\ccsdesc[500]{Software and its engineering~Domain specific languages}
\ccsdesc[300]{Hardware~Quantum computation}

\ccsdesc[500]{Software and its engineering~Domain specific languages}
\keywords{Quantum Programs, Program Synthesis, SMT solvers}  %% \keywords are mandatory in final camera-ready submission

%% environment and commands, and keywords command.
\maketitle
% \large
 \label{sec:intro}
\section{INTRODUCTION}

Quantum programming is a key step in enabling the various application of quantum computing such as factorization, simulation of physics and optimization. However, programming quantum computers is hard due to unintuitive quantum mechanics.
To help ease the programming of quantum computers, circuit synthesis techniques have been proposed to automatically generate quantum circuits~\cite{shende2006synthesis, de2020methods, amy2013meet, saeedi2011synthesis, kitaev1997quantum, younis2021qfast,kang2023modular}. 

Unfortunately, these synthesis frameworks underperform when the number of qubits of the quantum circuit to synthesize is as large as 5. For example, QFAST~\cite{younis2021qfast}, a recent quantum circuit synthesis framework, can only synthesize QFT and adder circuit up to 5 qubits. And QSyn~\cite{kang2023modular}, a quantum circuit synthesis method based on user-supplied components, needs an average time of 687.5 seconds for solving 4-qubit problems and fails to synthesize a 6-qubit circuit within one hour. These methods cannot scale with the rapid development of qubit numbers in hardware, with more than 1000 qubits by the end of 2023 estimated by IBM~\cite{ibm_roadmap}. 
Moreover, synthesizing a quantum circuit will even fail at the start because 
it is impossible to write the exponential-sized matrix specifying the synthesis goal. For example, QSyn requires 16 input-output state vector pairs as the specification for a 4-qubit Toffoli circuit. 
Additionally, the circuit generated by the synthesizer is hard to understand by humans, which prevents potential human customization to the circuit after synthesis.

\begin{figure}[h]
    \includegraphics[width=\textwidth]{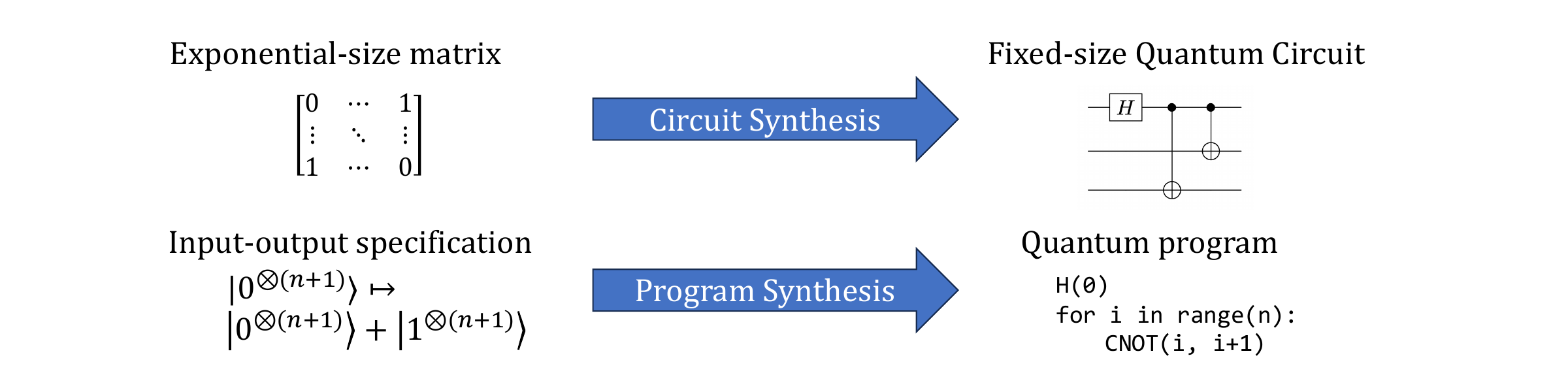}
    \caption{ Circuit synthesis vs. program synthesis. Circuit synthesis takes an exponential-sized matrix or state vector pairs as input and generates a fixed-size circuit,
    while program synthesis takes in an input-output specification and generates a program denoting a family of circuits for any input size.}
    \label{fig:compare}
\end{figure}

In this work, we propose \name, the first synthesis framework for inductive quantum programs. In contrast to previous frameworks focusing on circuits, the synthesis target of \name
is \emph{inductively-defined families of quantum circuits} without mid-circuit measurements (i.e. unitaries).
\name can exploit the inductive structure of quantum programs 
and, compared to previous circuit synthesis methods, 1) generate quantum circuits with an arbitrary number of qubits, 2) allow user to use a single input-output style specification for synthesis, and 3) produces more readable and structured quantum program code that are easy to customize by human, as illustrated by Figure~\ref{fig:compare}.

To enable the synthesis of inductive quantum programs, there are three challenges.
First, the synthesis framework requires a representation of the specification of quantum programs. Previous methods such as quantum Hoare triple~\cite{ying2012floyd}, path-sum~\cite{amy2018towards}, and tree automaton~\cite{chen2023automata} are all defined on fixed-dimension quantum systems and thus cannot be directly applied. Naive extensions of these representations with a variable qubit size do not work 
because symbolically representing matrices and automata is hard, limiting their usage for synthesis and verification.
We introduce a specification language in \name, which enables intuitive input-output style specification for quantum programs and supports all path-sum quantum states with an arbitrary number of qubits. We also propose the \textit{hypothesis-amplitude} ($h-\alpha$) specification that uses two functions to specify quantum programs. The specification written in \name-spec language will be compiled into $h-\alpha$ specification to use in the verification step, which avoids SMT-unfriendly matrices (or graphical representations like automata).

The second challenge is to find a subset of quantum programs that is both expressive enough and efficient to synthesize. Previous programming languages for verifying quantum programs are either low-level circuit languages that do not support inductive structure (e.g., SQIR~\cite{hietala2021voqc}) or high-level languages that do not have a detailed structure of unitaries (e.g., Quantum-while language used in Quantum Hoare Logic~\cite{ying2012floyd}). 
We design the Inductive-SQIR (ISQIR) language, an extension of SQIR that supports inductively defined quantum programs. We also define a Hoare-type verification logic that verifies an ISQIR program with respect to an $h-\alpha$ specification.

Finally, the synthesizer needs automated verification of inductive quantum programs. This requires reasoning about matrices, complex numbers, and formulas with a non-fixed number of terms, all of which have very limited support in SMT solvers.
To solve this challenge, we define \emph{parameterized path-sum amplitudes}, which, together with a sparsity constraint, can be efficiently encoded in SMT solvers. We further show that this set of expressions is expressive enough to specify and synthesize many quantum programs. 

We evaluate \name on \numbenchmark{} inductive quantum programs including state preparation, arithmetic and textbook quantum algorithm procedures. We showcase that \name can synthesize practical quantum programs including quantum adders~\cite{feynman1985quantum,cuccaro2004new}, a quantum subtractor, the eigenvalue inversion for HHL~\cite{lloyd2010quantum} algorithm, quantum teleportation~\cite{bennett1993teleporting} and Quantum Fourier Transform~\cite{coppersmith2002approximate}.  All synthesis processes succeed in 5 minutes, while many of the programs cannot be synthesized by previous methods when the number of qubits is larger than 6. A further investigation of synthesized programs shows that \name successfully captures the inductive structure of the targeting problem and produces better programs than human-written programs in Qiskit.

\paragraph{Contributions} Our contributions in this paper are multi-folded.

\begin{itemize}
    \item We propose \name, the first synthesis framework for inductively-defined quantum unitary circuit family.
    \item We introduce the \name-spec language that enables input-output style specification for inductive quantum programs, and the hypothesis-amplitude ($h-\alpha$) specification for scalable verification of quantum programs.
    \item We develop the syntax and the semantics of the inductive \sqir (\sqirs) language that supports recursively defined families of quantum unitary circuits, and a Hoare-type logic for proving the correctness of \sqirs program, with an $h-\alpha$ specification as predicates.
    \item We design Parameterized path-sum amplitude (\amplfunc) function, which leads to the efficient encoding of the verification process into SMT instances.
    \item We evaluate \name{} with a benchmark of \numbenchmark{} quantum programs, and show that \name{} is able to synthesize practical quantum programs. 
\end{itemize}

\section{QUANTUM PRELIMINARIES} \label{sec:qprelim}

In this section, we introduce the background knowledge about quantum program.
We recommend readers to refer \citet{nielsen_chuang_2010} for more details about quantum computing.
\subsection{Quantum States}
A quantum state consists of one or more \emph{quantum bits}. A quantum bit (or \emph{qubit}) can be expressed as a two dimensional vector $\begin{psmallmatrix} \alpha \\ \beta \end{psmallmatrix}$ such that $|\alpha|^2 + |\beta|^2 = 1$.
The $\alpha$ and $\beta$ are called \emph{amplitudes}.   We frequently write this vector as $\alpha\ket{0} + \beta\ket{1}$ where $\ket{0} = \begin{psmallmatrix} 1 \\ 0 \end{psmallmatrix}$ and $\ket{1} = \begin{psmallmatrix} 0 \\ 1 \end{psmallmatrix}$ are \emph{basis states}. A state written $\ket{\phi}$ is called a ket, following Dirac’s notation. When both $\alpha$ and $\beta$ are non-zero, we can think of the qubit as being ``both 0 and 1 at once,'' a.k.a. a \emph{superposition}. For example, $\frac{1}{\sqrt{2}}(\ket{0} + \ket{1})$ is an equal superposition of $\ket{0}$ and $\ket{1}$. A qubit is only in superposition until it is \emph{measured}, at which point the outcome will be $0$ with probability $|\alpha|^2$ and $1$ with probability $|\beta|^2$.  \par
A quantum state with $n$ qubits is represented as vector of length $2^n$. We can join multiple qubits together by means of the \emph{tensor product} ($\otimes$) from linear algebra. For convenience, we write $\ket{x_0} \otimes \ket{x_1} \otimes \cdots \otimes \ket{x_m}$ as $\ket{x_0x_1...x_m}$ for $x_i \in \{0,1\}, 0\le i \le m$; we may also write $\ket{k}$ where $k \in \mathbb{N}$ is the decimal interpretation of bits $\ket{x_0x_1...x_m}$. For example, a 2-qubit state is represented as a $2^2=4$ length vector where each component corresponds to (the square root of) the probability of measuring $\ket{00}$, $\ket{01}$, $\ket{10}$, and $\ket{11}$, respectively. We may also write these four kets as $\ket{0},\ket{1},\ket{2},\ket{3}$.
 Sometimes a multi-qubit state cannot be expressed as the tensor product of individual qubits; such states are called \emph{entangled}. One example is the state $\frac{1}{\sqrt{2}}(\ket{00} + \ket{11})$, known as a \emph{Bell pair}.
 
\subsection{Quantum Programs}
\begin{wrapfigure}{R}{.18\textwidth}
$\begin{pmatrix} 
1 & 0 & 0 & 0 \\ 
0 & 1 & 0 & 0 \\ 
0 & 0 & 0 & 1 \\ 
0 & 0 & 1 & 0 
\end{pmatrix}$
\end{wrapfigure}
Quantum programs are composed of a series of \emph{quantum operations}, each of which acts on a subset of qubits. Quantum operations can be expressed as matrices, and their application to a state is expressed as matrix multiplication. For example, the \emph{Hadamard} operator $H$ on one qubit is expressed as a matrix 
$\frac{1}{\sqrt{2}}\begin{psmallmatrix} 1 & 1 \\ 1 & -1 \end{psmallmatrix}$.
Applying $H$ to state $\ket{0}$ yields state $ \frac{1}{\sqrt{2}}\ket{0}+ \frac{1}{\sqrt{2}}\ket{1}$. $n$-qubit operators are represented as $2^n \times 2^n$ matrices. For example, the $\mathit{CNOT}$ operator over two qubits is expressed as the $2^2\times 2^2$ matrix shown at the right. \par
\begin{figure}[h]
    \captionsetup[subfigure]{justification=centering}
\begin{minipage}[b]{.25\textwidth}
\centering
  \small
  \centerline{
  \Qcircuit @C=0.5em @R=0.5em {
    \lstick{\ket{0}} & \gate{H} & \ctrl{1} & \qw & \qw \\
    \lstick{\ket{0}} & \qw & \targ & \ctrl{1} & \qw \\
    \lstick{\ket{0}} & \qw & \qw & \targ & \qw
    }  
  }

\subcaption{Quantum circuit}
\end{minipage}
\begin{minipage}[b]{.2\textwidth}
\begin{coq}
H 0; 
CNOT 0 1; 
CNOT 1 2;
\end{coq}
\subcaption{Unitary \sqir}
\end{minipage}

~
\begin{minipage}[b]{0.45\textwidth}
\centering
$(I \otimes CNOT) \times (CNOT \otimes I) \times (H \otimes I \otimes I)$
\subcaption{Matrix expression}
\end{minipage}
\caption{Example quantum program: 3-qubit GHZ state preparation.}
\label{fig:circuit-example}
\end{figure}
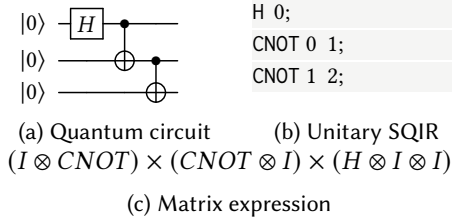

In the standard presentation, quantum programs are expressed as \emph{circuits}, as shown in \Cref{fig:circuit-example}(a). In these circuits, each horizontal wire represents a qubit and boxes on these wires indicate quantum operations, or \emph{gates}. The circuit in \Cref{fig:circuit-example}(a) has three qubits and three gates: the \emph{Hadamard} (\texttt{H}) gate and two \emph{controlled-not} (\texttt{CNOT}) gates. The semantics of a gate is a \emph{unitary matrix}. Applying a gate to a state is tantamount to multiplying the state vector by the gate's matrix. The matrix corresponding to the circuit in \Cref{fig:circuit-example}(a) is shown in \Cref{fig:circuit-example}(c), where $I$ is the $2 \times 2$ identity matrix. 
\subsection{Unitary \sqir}
\sqir\cite{hietala2021voqc} is a simple quantum language embedded in the Coq proof assistant. 
\sqir's \emph{unitary fragment} is a sub-language for expressing programs consisting of unitary gates. 
\paragraph{Syntax} A unitary \sqir program $P$ is a sequence of applications of gates $G$ to qubits $q$:
\begin{align*}
    P:=P_1;P_2 \mid G~q \mid G~q_1~q_2 \mid G~q_1~q_2~q_3.
\end{align*}
Qubits are referred to by natural numbers that index into a global register. A
\sqir program is parameterized by a unitary gate set $g$ (from which $G$ is drawn) and the size $n$ of the global register (i.e., the number of available qubits). In Coq, a unitary \sqir program \coqe{U} hence has type \coqe{ucom g n}.
$U$ has type \coqe{ucom }$g~n$, where $g$ identifies the gate set and $n$ is the size of the global register.

\begin{wrapfigure}{L}{.43\textwidth}
\begin{coq}
Fixpoint ghz (n : nat) : ucom g (n + 1) :=
  match n with
    | 0 => H 0
    | S n' => ghz n'; CNOT n' n
  end.
\end{coq}
\end{wrapfigure}
The Coq function \coqe{ghz} on the left recursively constructs a \sqir program, which prepares the $n+1$-qubit GHZ state.
When $n=0$, the program applies the Hadamard gate $H$ to qubit $0$. Otherwise, \coqe{ghz} calls itself recursively with input $n-1$ and then applies $\mathit{CNOT}$ to qubits $q_{n-1}, q_n$. 
For example, \coqe{ghz 2} generates the circuit in \Cref{fig:circuit-example}(a).
\paragraph{Semantics}
\begin{figure}
  \centering
  \begin{gather*}
    \sem{P_1;~P_2}_d=~\sem{P_2}_d \times \sem{P_1}_d; \\
    \sem{G~ q_1\cdots q_i}_d=\begin{cases}
                          apply_i(G,~q_1,...,~q_i,~d) &\text{well-typed} \\
                          0_{2^{d}} &\text{otherwise}
                          \end{cases},~ i=1,2,3.
  \end{gather*}
  
\caption{Semantics of unitary \sqir programs, assuming a global register of dimension $d$. The $apply_k$ function maps a gate name to its corresponding unitary matrix and extends the intended operation to the given dimension by applying an identity operation on every other qubit in the system.}
  \label{fig:sqire-semantics}
\end{figure}
The semantics of unitary \sqir is shown in \Cref{fig:sqire-semantics}. A program $P$ is well-typed if every gate's index arguments are within the bounds of the global register and no index is repeated. The program's semantics follows from the composition of the matrices that correspond to each of the applications of its unitary gates. A gate application's matrix needs to apply the identity operation to the qubits not being operated on. This is the purpose of using $apply_1,apply_2$ and $apply_3$ 
For example, $apply_1(G_u,~q_1,~d) = I_{2^q} \otimes u \otimes I_{2^{(d - q - 1)}}$ where $u$ is the matrix interpretation of the gate $G_u$ and $I_k$ is the $k \times k$ identity matrix.

Suppose that $M_1$ and $M_2$ are the matrices corresponding to unitary gates $P_1$ and $P_2$, which we want to apply to a quantum state vector $\ket{\psi}$. Matrix multiplication is associative, so $M_2(M_1 \ket{\psi})$ is equivalent to $(M_2 M_1) \ket{\psi}$. Moreover, multiplying two unitary matrices yields a unitary matrix. As such, the semantics of \sqir program $P_1;~ P_2$ is naturally described by the unitary matrix $M_2 M_1$. \Cref{fig:circuit-example}(b) shows an example unitary \sqir program for the circuit in \Cref{fig:circuit-example}(a).

\subsection{Path-sum Representation}
\label{sec:path-sum}
 \textit{Path-sum}, proposed by recent works on quantum program verification~\cite{amy2018towards,qbricks2020deductive}, is a representation for describing quantum states based on Feynman’s \textit{path integral} formalism of quantum mechanics, which is widely applied to circuit simulation~\cite{koh2017computing,bravyi2016improved} and optimization~\cite{amy2018controlled,amy2014polynomial,amy2019tcount}. 
The idea of this formalism is to describe a quantum state's amplitude by an integral over all paths leading to that state. In practice, a discrete sum-over-path technique rather than integral is typically used ~\cite{bacon2008analyzing,dawson2005solovay,montanaro2017quantum,koh2017computing,amy2018towards,amy2014polynomial,amy2019tcount,bravyi2016improved,qbricks2020deductive}. We can describe a sum-over-path abstractly as a discrete set of paths $T\in \mathbb{Z}_2^m$, together with an amplitude
function $\psi$ and a state transformation $f$, both depending on specific path $y$, to represent a unitary $U$: 
\begin{align*}
    U:\ket{x} \mapsto \sum_{y\in T} \psi(x,y) \ket{f(x,y)}. 
\end{align*}
All representations based on the sum-over-path in these previous works\cite{amy2018towards,qbricks2020deductive,bacon2008analyzing,dawson2005solovay,montanaro2017quantum,koh2017computing} share in common that the amplitude $\psi(x,y)$ of all possible paths have the same \textit{magnitude} and only the \textit{phase}s are different. These forms of the same \textit{magnitude} but different \textit{phase} are so typical in many quantum algorithms that they can be used as a succinct representation in the verification, which makes them useful for synthesis purposes.

\section{OVERVIEW} \label{sec:overview}

\begin{figure}[h]
    \centering
    \includegraphics[width=0.8\textwidth]{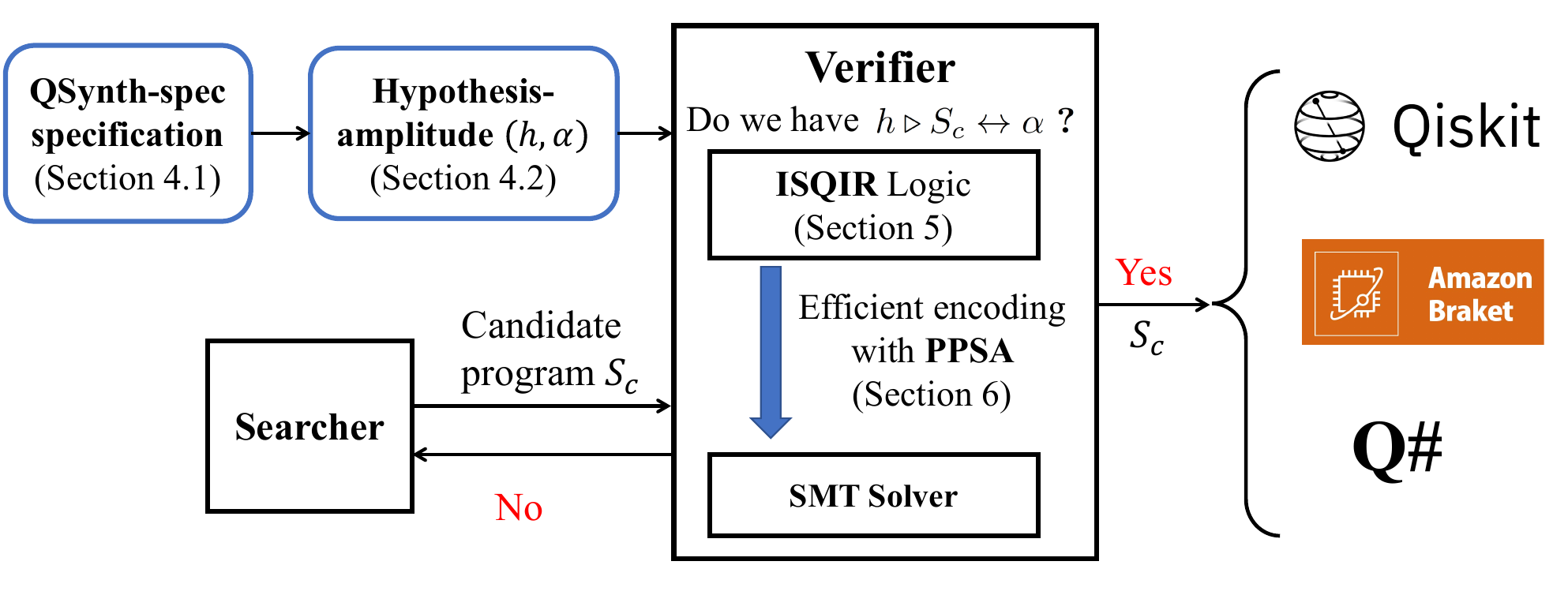}
    \vspace{-10pt}
    \caption{\name Overview.}
    \label{fig:overview}
\end{figure}

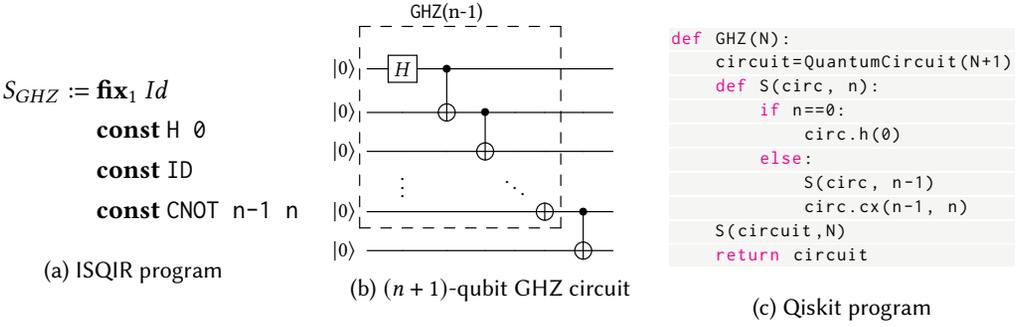
\begin{figure}
    \begin{minipage}[p]{0.25\textwidth}
    {
    \begin{align*} 
S_{GHZ}:=~& \textbf{fix}_1~Id\\
    & \const{\texttt{H 0}}\\
    &\const{\texttt{ID}}\\
    &\const{\texttt{CNOT n-1 n}}
\end{align*}}%
\subcaption{\sqirs program}
    \end{minipage}
    ~
    \begin{minipage}[p]{.33\textwidth}
    \centerline{
\scalebox{0.8}{
\Qcircuit @C=1.0em @R=1.0em {
& & \text{\texttt{GHZ}(n-1)}\\
&&&&&&&\\
    \lstick{\ket{0}} & \gate{H} & \ctrl{1} & \qw & \qw & \qw & \qw &\qw\\
    \lstick{\ket{0}} & \qw & \targ & \ctrl{1} & \qw & \qw & \qw &\qw\\
    \lstick{\ket{0} } & \qw & \qw & \targ &\qw & \qw & \qw & \qw & \\
      & \vdots & &&\ddots & & & \\
    \lstick{\ket{0}} &\qw& \qw & \qw & \qw & \targ & \ctrl{1} & \qw  \gategroup{2}{1}{7}{6}{0.7em}{--}\\
    \lstick{\ket{0}} &\qw & \qw &\qw & \qw & \qw & \targ & \qw \\
}
}}
\subcaption{$(n+1)$-qubit GHZ circuit}
~
\end{minipage}
    \begin{minipage}[p]{.33\textwidth}
\begin{lstlisting}[language=Python]
def GHZ(N):
    circuit=QuantumCircuit(N+1)
    def S(circ, n):
        if n==0:
            circ.h(0)
        else:
            S(circ, n-1)
            circ.cx(n-1, n)
    S(circuit,N)
    return circuit
\end{lstlisting}
\subcaption{Qiskit program}
\label{fig:qiskit}
\end{minipage}

    \caption{$(n+1)$-qubit GHZ state preparation programs in different programming languages. (a) \sqirs program $S_{GHZ}$. (b) The quantum circuit represented by $S_{GHZ}$. (c) Qiskit function compiled from $S_{GHZ}$. Statement \texttt{circ.cx} in Qiskit means appending a \texttt{CNOT} gate to the circuit. Qiskit's semantic requires the program always use class \texttt{QuantumCircuit(n)} to initialize a circuit. So \name compiler will wrap program $S$ with an outside function \texttt{GHZ} to initialize the circuit. 
    }
    \label{fig:compile-ex}
    \vspace{-10pt}
\end{figure}

The workflow of \name is shown in Fig~\ref{fig:overview}. 
First, the user writes an input-output style specification of the program to be synthesized 
using the \name-spec language.
Then, \name will translate the specification into a hypothesis-amplitude (h-$\alpha$) pair for later verification.
Meanwhile, \name will invoke a syntax-guided program searcher~\cite{sumit2017foundations} to 
generate all possible candidate programs written in the ISQIR language within a given search space.
For each candidate ISQIR program, \name will use the unitary ISQIR logic to verify 
whether the program satisfies the (h-$\alpha$) specification. 
The verification process will be done in an SMT solver in which all the numbers are encoded
in the form of parameterized path-sum amplitudes (PPSA).
If the verification process succeeds, a correct program is synthesized and the program can be translated 
by \name's ISQIR compiler into commercial quantum programming languages including Qiskit, Q\# and Braket,
as shown in Figure~\ref{fig:compile-ex}.
If the verification fails, the searcher will try the next candidate program.

Next, we will walk through \name's components using the synthesis of the GHZ state preparation program as an example.

\paragraph{Target Program}
An $N$-qubit ($N\in Z^+$) Greenberger-Horne-Zeilinger state (GHZ state) is an entangled quantum state given by 
\begin{align*} \label{eq:GHZ}
    \ket{GHZ}_N = \frac{1}{\sqrt{2}} ( \ket{0}^{\otimes N} + \ket{1}^{\otimes N}),
\end{align*}
which was first studied by~\citet{greenberger1989going} and is widely used in quantum information, e.g., \cite{xia2006quantum,zhong2006controlled, hillery1999quantum,liao2014dynamic}.
Preparing the $N$-qubit GHZ state from $\ket{0}^{\otimes N}$ for any $N$ is a natural task for program synthesis. 

Namely, we hope to synthesize an \sqirs program $S$ such that $\sem{\ssem{S}(n)}$, the instantiation of $S$ with $n\in \mathbb{N}$ where $\mathbb{N}$ denotes the natural number in the following paper, is a unitary that transfers state $\ket{0}^{n+1}$ to state $\ket{GHZ}_{n+1}$ for any $n\geq 0$. 

\paragraph{Target Specification} Our first challenge is how to specify our target program, which generates $\ket{GHZ}_{n+1}$ from $\ket{0}^{n+1}$. The difficulty is two-folded: (1) we want a specification for any input $n$, which excludes any existing specification methods for fixed dimensions (e.g., Quantum Hoare triples); (2) the specification should be as succinct as possible. 
Then, a direct matrix representation that might require $2^{n+1} \times 2^{n+1}$ is less desirable. 

In \name{}, the specification is given in an input-output manner using the \name-spec language in the form of $\text{GHZ} : \ket{0_{n+1}} \mapsto \ket{0_{n+1}} \uplus \ket{(2^{n+1}-1)_{n+1}}$, where $\ket{a} \uplus \ket{b}$ means the normalized sum $(\ket{a}+\ket{b})/\sqrt{2}$.
Then, the input-output style specification is compiled into the following \emph{hypothesis-amplitude $(h-\alpha)$ specification} (Definition~\ref{def:judge}) for later verification. An instantiation of $(h-\alpha)$ to the GHZ target program is given as follows: 
\begin{align}
    h&:= \{(n,x,y)~|~x=0\},\qquad
        \alpha_{GHZ}(n,x,y) := \frac{1}{\sqrt{2}}\bvalue{y=0\lor y=2^{n+1}-1},
\end{align}
where the term $\bvalue{b}$ in $\alpha_{GHZ}$ is a \{0,1\}-valued function that returns 1 if the Boolean expression $b$ is True and 0 otherwise. 

Intuitively, the hypothesis function $h$ specifies the interesting input to the program, the desired program's behaviour which is specified by the amplitude function $\alpha$. 
Precisely, given input $x$, the amplitude $\bra{y}\sem{\ssem{S}(n)}\ket{x}$ of basis $y$ on the input $x$ for a desired program $S$ is given by $\alpha(n,x,y)$, where $x,y$ are bit strings.

In the GHZ example, we are only interested in input $\ket{0}^{n+1}$, which leads to a trivial $h$ containing only $x=0$ in (\ref{eq:GHZ}). 
The output state, which is $\frac{1}{\sqrt{2}}(\ket{0}^{n+1} + \ket{1}^{n+1})$, corresponds to an amplitude function $\alpha(n,x,y)$ with only non-zero value $\frac{1}{\sqrt{2}}$ on either $y=0$ (referring to $\ket{0}^{n+1}$) or $y=2^{n+1}-1$ (referring to $\ket{1}^{n+1}$), which explains (\ref{eq:GHZ}). 

Our hypothesis-amplitude specification is arguably as natural as the common classical specifications that describe the desired input-output relationship, except that one could have many such input-output pairs (i.e., \emph{superposition}) with potential complex amplitudes, in the quantum setting, which requires an explicit use of our amplitude function $\alpha(n,x,y)$.

\begin{figure}
    \centering
    \includegraphics[width=0.95\textwidth]{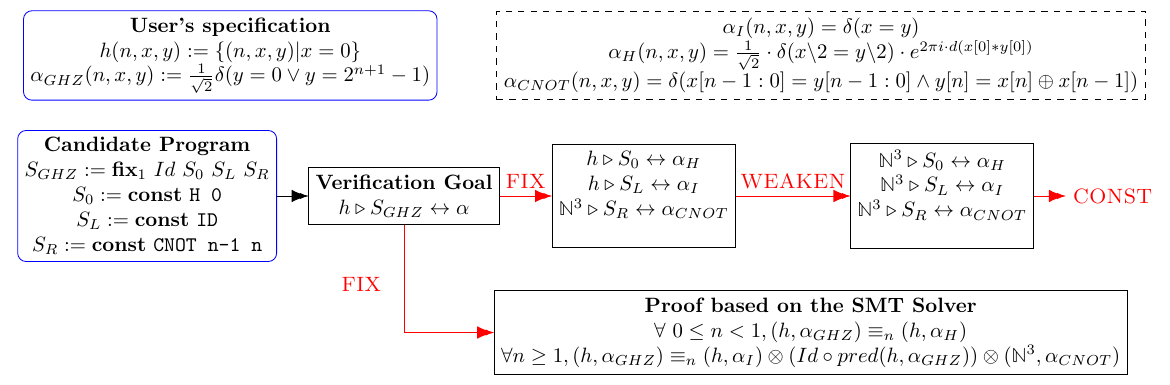}
    \caption{An example of synthesizing $n+1$-qubit GHZ state preparation program. }
    \label{fig:synthesis-ex}
    \vspace{-10pt}
\end{figure}

\paragraph{Verification of Candidate Programs}
Assume the \name searcher has identified a candidate $S_{GHZ}$, same as Fig~\ref{fig:compile-ex} (a), on the left of Fig~\ref{fig:synthesis-ex}. The program $S_{GHZ}$ is constructed by a \textsc{FIX} syntax with subprograms $S_0 = \texttt{H 0}$, $S_L = \texttt{ID}$ and $S_R = \texttt{CNOT n-1 n}$, which is a recursive program with the base case $S_{GHZ}(0) := S_0$ and the inductive case, $S_{GHZ}(n) := S_L(n); S_{GHZ}(n-1); S_R(n)$. This \sqirs program is equivalent to the recursive Qiskit program in Figure~\ref{fig:qiskit}. The \textsc{FIX} syntax, similar to the fixpoint in Coq, enables inductive structures in \sqirs programs.

\name verifier leverages the newly developed unitary \sqirs logic (Section~\ref{sec:logic}) to verify the goal judgement $\interpret{h}{S_{GHZ}}{\alpha_{GHZ}}$, which basically states that candidate program $S_{GHZ}$ satisfies the $(h, \alpha)$ specification. \name verifier recursively applies the logic rules to split the judgement of $h,\alpha$ for larger programs into that of smaller programs. The side conditions are checked by SMT solvers, and the $h,\alpha$ judgement for constant SQIR program for quantum gates that are independent of $n$ are directly computed.

In the GHZ example, \name verifier first uses the \textsc{FIX} rule to split the goal judgement into two parts. The first part is two formulas (right bottom of Fig~\ref{fig:synthesis-ex}) to be checked by the SMT solver, with details elaborated on later. 
The second part is three judgements for $S_L, S_R, S_0$ (right hand of the verification goal box in Fig~\ref{fig:synthesis-ex}). After applying the \textsc{FIX} rule \name uses the \textsc{WEAKEN} rule to adjust these judgements into an appropriate format. 
Since $S_0, S_L, S_R$ are simple programs (formally called const SQIR programs), their corresponding hypothesis-amplitude specifications (i.e. $\alpha_I, \alpha_H, \alpha_{CNOT}$ on the upper right corner of 
Fig~\ref{fig:synthesis-ex}) are predefined in \name and can be verified directly using the \textsc{CONST} rule.
$\alpha_I$ returns 1 if $x=y$ else 0, indicating the identity matrix.
 $\alpha_H$ represents the matrix of the program $\const{H 0}$, i.e., $\frac{1}{\sqrt{2}}\begin{psmallmatrix} 1 & 1 \\ 1 &-1 \end{psmallmatrix} \otimes I^n$. The expression $e^{2\pi i\cdot \frac{x[0]*y[0]}{2}}$ in $\alpha_H$ returns $-1$ if $x[0]=1,y[0]=1$ and returns 1 otherwise, indicating the matrix $\frac{1}{\sqrt{2}}\begin{psmallmatrix} 1 & 1 \\ 1 &-1 \end{psmallmatrix}$; the expression $\delta(x\backslash 2=y\backslash 2)$, where $x\backslash 2$ and $y\backslash 2$ are integer division (e.g. $10\backslash 3=3$), returns 1 if $x,y$ only have the lowest bit difference and 0 otherwise, indicating the matrix $I^n$ on qubits $q_1,...,q_n$. $\alpha_{CNOT}$ will be discussed later.
 After this verification step, the synthesis terminates and \name compiles $S_{GHZ}$ into programs as shown in Fig~\ref{fig:compile-ex}.

\paragraph{SMT solving with PPSA}
We continue with the two formulas in the bottom right corner box in Fig~\ref{fig:synthesis-ex} generated by the FIX rule and aim to verify them by SMT solvers. 

The first formula indicates the \emph{base} case is correct. The relation $\equiv_n$ indicates the equivalence between two hypothesis-amplitude specifications given specified $n$. In this GHZ example, the first formula becomes:
\begin{align*}
\forall x~y\in \nat, x=0 \rightarrow \alpha_{GHZ}(0,x,y) = \alpha_H(0,x,y).
\end{align*}
Note that the amplitude functions $\alpha_{GHZ}$ or $\alpha_H$ are generally complex-valued, the automatic equivalence check of which are not generally supported by any existing tool. 

Inspired by recent work on quantum program verification~\cite{amy2018towards,qbricks2020deductive}, \name restricts $\alpha$ into a succinct path-sum representation yet with rich enough expressiveness (elaborated on in Section~\ref{sec:encode-alpha}), called the parameterized path-sum amplitude (PPSA) in Definition~\ref{def:ppsa}. 
In PPSA representation, the non-zero amplitudes over $x,y$ share the same magnitude, which depends on $n$, but can have different phases. 
Thus, instead of representing a general amplitude function $\alpha(n,x,y)$,  it suffices to represent $\alpha(n,x,y)$ by components. For instance, $\alpha_H, \alpha_{GHZ}$ defined in Fig~\ref{fig:synthesis-ex}, can be described with three components: 
\begin{itemize} [leftmargin=5mm]
    \item A fraction expression indicating the amplitude: $\frac{1}{\sqrt{2}}$ for both $\alpha_{GHZ}, \alpha_{H}$.
    \item A Boolean expression indicating the non-zero value: $\bvalue{y=0\lor y=2^{n+1}-1}$ for $\alpha_{GHZ}$ and $\bvalue{x\backslash 2=y\backslash 2}$ for $\alpha_H$.
    \item An expression indicating the phase: $e^{2\pi i \cdot 0}$ for $\alpha_{GHZ}$ and $e^{2\pi i \cdot \frac{x[0]*y[0]}{2}}$ for $\alpha_H$, where $x[0]$ means the lowest (i.e. 0th) bit of $x$'s binary representation (e.g. $6[0]=(110)_b=0$).
\end{itemize}
A direct substitution of $\alpha_H, \alpha_{GHZ}$ would require \name, or SMT solvers, to verify 
\begin{align*}
   &\forall x~y\in \nat, x=0 \rightarrow   \frac{1}{\sqrt{2}}\bvalue{y=0\lor y=1} = \frac{1}{\sqrt{2}}\bvalue{x\backslash 2=y\backslash 2}\cdot e^{2\pi i\cdot \frac{x[0]*y[0]}{2}},
\end{align*}
which is infeasible. Using PPSA, one can equivalently verify the following by SMT solvers:
\begin{align*}
 \forall x~y \in \mathbb{N}, x=0 \rightarrow  \frac{1}{\sqrt{2}}=\frac{1}{\sqrt{2}}  \land  \bvalue{y=0\lor y=1} = \bvalue{x\backslash 2=y\backslash 2}
   \land \frac{x[0]*y[0]}{2} = 0.   
\end{align*}

The second formula concerns the correctness of the \emph{induction} case. The function $pred(h,\alpha_{GHZ})$ generates the hypothesis-amplitude specification for the recursive call (i.e. $S_{GHZ}(n-1)$), and $(h_1, \alpha_1) \otimes (h_2, \alpha_2)$ calculates the one when composing two \sqirs programs together, both formally in Definition~\ref{def:calculation}. 
In this example, the composition with $(h, \alpha_I)$ is trivial since $\alpha_I$ represents an identity matrix, and the second formula becomes 
\begin{gather}
    \forall n\ge 1, \forall x~y \in \mathbb{N}, (h, \alpha_{GHZ}) \equiv_n (h, \alpha'), \nonumber\\ \text{where }\alpha'(n,x,y) = \sum_{z\in \nat} \alpha_{GHZ}(n-1,x,z)\alpha_{CNOT}(n,z,y),\label{equ:compose}\\
    \alpha_{CNOT}(n,x,y)=\bvalue{x[n-1:0]=y[n-1:0] \land y[n]=x[n]\oplus x[n-1]}.\nonumber
\end{gather}
$\alpha_{CNOT}$ is designed to represents $S_R$: the term $x[n-1:0]=y[n-1:0]$ indicates the identity map $\ket{q_0...q_{n-1}} \mapsto \ket{q_0...q_{n-1}}$; the term $y[n]=x[n]\oplus x[n-1]$ indicates the map $\ket{q_{n-1}}\ket{q_n}\mapsto \ket{q_{n-1}}\ket{q_{n-1}\oplus q_n}$. Their combination indicates $S_R$'s map, i.e., $\ket{q_0...q_{n-1}}\ket{q_n}\mapsto \ket{q_0...q_{n-1}}\ket{q_{n-1}\oplus q_n}$.
% $x[n-1:0]$ equals to $(x \text{ mod } 2^n)$ mathematically when $n\ge 1$.

There is another major challenge to verify (\ref{equ:compose}) by SMT solvers due to the infinite summation over $z\in \mathbb{N}$, which comes from the composition of two amplitude specifications (details in Section~\ref{sec:formulation}).  
As a result, $\alpha'$ could have an unbounded number of terms, which makes it infeasible for any SMT solver. 

To circumvent this general difficulty, we introduce a \emph{sparsity} constraint, which restricts the number of non-zero points with any fixed $x$ or $y$ to be constant (Definition~\ref{def:sparsity}), and prove that the composition of two amplitude functions will have finite terms if \textit{one of} the composed function is sparse. We also prove that all quantum gates applied on a constant number of qubits (e.g. all SQIR programs) have a sparse amplitude function. We only apply this sparsity constraint to SQIR programs predefined in \name, and \textit{non-sparse} functions can be generated as synthesis specifications.
However, the use of the FIX statement could generate non-sparse amplitude functions. 

As a result, we only allow one use of the FIX statement in our synthesis, as otherwise we could risk composing two non-sparse amplitude functions that would lead to an infinite sum. 
The specification for the target program, however, could be non-sparse, as we won't need to compose the target programs with others. 

% 
%As long as the ISQIR program has only one FIX statement, the specification function is only a non-sparse one during the composition and there will be no infinite sums.

In the GHZ example, $\alpha_{CNOT}$ is sparse and we have
%Fortunately, we have the following property for $\alpha_{CNOT}(n,z,y)$
\begin{align*}
    \forall n, y\in\nat, \quad \alpha_{CNOT}(n,z,y)\neq 0 \leftrightarrow z=y\oplus (y[n-1]\text{<<}n). 
\end{align*}
Here the expression $y\oplus (y[n-1]\text{<<}n)$ sets the $(n+1)$th bit of $y$ (i.e. $y[n]$) to $y[n]\oplus y[n-1]$ and keeps $y[n-1:0]$ unchanged. 
%This sparsity property says that the \sqirs program that $\alpha_{CNOT}$ represents, $\const{\texttt{CNOT n-1 n}}$, only modifies the state of a constant number of qubits (i.e. qubit $n$). 
% We say any amplitude function with such property is \textit{sparse}, which is formally defined in Section~\ref{sec:encoding}. Any quantum gates applied on constant number of qubits (e.g. Clifford-T gate set, Toffoli gate) have a \textit{sparse} amplitude function. 
Let $z=y\oplus (y[n-1]\text{<<}n)$. 
With this sparsity of $\alpha_{CNOT}$, we can simplify the formula (\ref{equ:compose}) to
\begin{align*}
      &\forall n\ge 1, \forall x~y \in \mathbb{N}, (h, \alpha_{GHZ}) \equiv_n (h, \alpha'),    \\
    \text{ where } &\alpha'(n,x,y) =  \alpha_{GHZ}(n-1,x,z)\alpha_{CNOT}(n,z,y) =  \frac{1}{\sqrt{2}}\bvalue{z=0 \lor z= 2^{n}-1}. 
\end{align*}
%\name encodes the relation $\equiv_n$ in the same way for the first formula and generates the formula below.
% After a direct substitution of $\alpha_{GHZ}$, which happens to be a complex-value free function, 
% \begin{align*}
%  &\forall n\ge 1,  \forall x~y\in \nat, x=0 \rightarrow   \frac{1}{\sqrt{2}}\bvalue{y=0\lor y=2^{n+1}-1} = \frac{1}{\sqrt{2}}\bvalue{z=0 \lor z= 2^{n}-1}.
% %  & z=y\oplus (y[n-1]\text{<<}n)
% \end{align*}
With the PPSA representation, it suffices to verify the following SMT instance:
\begin{align*}
   \forall n\ge 1, \forall x~y \in \mathbb{N}, x=0 \rightarrow  \frac{1}{\sqrt{2}}=\frac{1}{\sqrt{2}} &~ \land ~ \bvalue{y=0\lor y=2^{n+1}-1} = \bvalue{z=0 \lor z= 2^{n}-1}.
\end{align*}

\paragraph{Organization}
In Section~\ref{sec:spec} we describe the \name's specification language and hypothesis-amplitude specification.
In Section~\ref{sec:formulation} we introduce the ISQIR programming language for the inductive quantum circuit family, 
and its associated Hoare-type logic,
In Section~\ref{sec:encoding} we introduce the PPSA encoding.
In Section~\ref{sec:case} we discuss the implementation and the evaluation of \name on a benchmark of \numbenchmark{} programs.
In Section~\ref{sec:relatedwork} we discuss the related work.
In Section~\ref{sec:future} we discuss the limitation of \name and future work.
In Section~\ref{sec:conclusion} we give the conclusion.

\section{SPECIFICATION}
In this section, we explain how specifications are provided and processed in \name{}. Users provide the input-output specification in the \name-spec language (defined in Section~\ref{sec:spec-lang}).
Then, the specification will be compiled into a hypothesis-amplitude pair~(defined in Section~\ref{sec:h-alpha}) as the predicate for later verification. We describe the compilation process in Section~\ref{sec:spec2ha}
\label{sec:spec}
\subsection{\name-spec Interface}
\label{sec:spec-lang}
To allow users to give the synthesis specification more intuitively, 
we design language \name-spec shown in Fig~\ref{fig:spec-lang}. Programmers provide the synthesis specification in \name-spec, and \name will compile it to the corresponding hypothesis-amplitude specification.

\begin{figure}[h]
    \centering\small
    $$
\begin{array}{lccllccl}
\textit{(Spec)} & \mathcal{S} &::= &\mathcal{I} \mapsto \mathcal{O}\\
\textit{(Input)} & \mathcal{I} &::= &\ket{c_l} \mid \ket{v[\ell]} \mid \mathcal{I}_1 \otimes \mathcal{I}_2\\
\textit{(Output)} & \mathcal{O} &::= &\multicolumn{5}{l}{\ket{E} \mid e^{i\cdot E} \cdot O' \mid \mathcal{O}_1 \otimes \mathcal{O}_2 \mid e^{i \cdot E_1} \ket{c_1} \uplus e^{i \cdot E_2} \ket{c_2}  \mid \biguplus_{z\in \{0,1\}^{l}} \delta(B)\cdot e^{-2\pi i\cdot E}\ket{z}}\\
\textit{(VarExp)} & E&::= &c_\ell\mid v \mid \text{\textbf{uop }} E' \mid E_1 \text{\textbf{ bop }} E_2 \\
\textit{(BoolExp)} & B &::= &E_1 \text{\textbf{ rop }} E_2 \mid B_1 \land B_2 \mid B_1 \lor B_2 \\
\textit{(Length)} & \ell &::= &c \mid \text{\texttt{n}} \mid \text{\textbf{uop }} \ell' \mid \ell_1 
\text{\textbf{ bop }} \ell_2\\

\textit{(Variable)} &  v & \in & \text{Variables} & \textit{(Constant)} &c&\in& \nat
\end{array}$$
    \caption{\name-spec syntax. \textbf{uop} and \textbf{bop} are common unary operators and binary operators (e.g. + - * /)}. \textbf{rop} are common relation operators (e.g. ==, !=, >, <). $\delta(B)$ is a function that returns 1 if $B$ is True and returns 0 otherwise. $\uplus$ means normalized summation.
    \label{fig:spec-lang}
\end{figure}

A \name-spec specification is given in the form $Input \mapsto Output$. $Input$ indicates the input quantum state to the desired unitary. It is a ket expression constructed by $l$-bit constant numbers $c_l$, a $l-$bit variable $v[l]$, or the tensor product of two quantum states. Programmers can arbitrarily declare variables in the $Input$. For example, consider the $Input$ specification for a $n$-bit quantum adder: $\ket{A}\ket{B}\ket{0}^n \mapsto \ket{A}\ket{B}\ket{A+B}$.
\begin{align}
\ket{0}\ket{A[n]}\ket{B[n]}\ket{0_n} \mapsto \ket{c_0}\ket{A}\ket{B}\ket{A+B}
\label{equ:adderspec}
\end{align}
 where $c_0$ is the carry bit of $A+B$. We simplify $\ket{\phi}\otimes \ket{\psi}$ as $\ket{\phi}\ket{\psi}$ and omit the length specification for the one-bit state $\ket{0}, \ket{1}$. This specification indicates the qubit $q_0$ is initialized as $\ket{0}$ to hold the carry bit; qubits $q_1\sim q_n$ and qubits $q_{n+1}\sim q_{2n}$ store two $n$-bit numbers $A[n], B[n]$; qubit $q_{2n+1}\sim q_{3n}$ are initialized to $\ket{0}^n$ to store the sum of $A[n], B[n]$.
 
$Output$ suggests the state transformed from $Input$ by the target unitary program. $Output$ can be constructed by a variable expression $E$, a state shifted by the phase $e^{iE}$, or the tensor product of two states. The variables that appear in $E$ are bound: they can only be a variable declared in $Input$ or the sum variable $y$ when it is in the sum scope. Output can also be constructed as the superposition state with the same magnitude but different phase by $e^{i \cdot E_1} \ket{c_1} \uplus e^{i \cdot E_2} \ket{c_2}$ and $\biguplus\nolimits_{z\in \{0,1\}^{l}} \delta(B)\cdot e^{-2\pi i\cdot E}\ket{z}$. For example, consider the specifications for $n$-qubit GHZ state program and QFT program. 
\begin{align}
 &\text{GHZ}_n : \ket{0_n} \mapsto \ket{0_n} \uplus \ket{(2^n-1)_n} \qquad
 \text{QFT}_n : \ket{x[n]} \mapsto \biguplus\nolimits_{y\in \{0,1\}^n} e^{-2\pi i\cdot \frac{xy}{2^n}}
 \label{equ:qftspec}
\end{align}
Programmers can omit the amplitude normalization term, which will be calculated by \name during the compilation from \name-spec to the hypothesis-amplitude specification.

\subsection{Hypothesis-amplitude Specification for Verification}
\label{sec:h-alpha}
A critical component of program synthesis is the ability of expressing desired properties of the target programs, usually called specifications. 
The pre and post conditions of programs in typical Hoare triples provide a natural approach to express these input-output specifications. 
Quantum Hoare triples~\cite{ying2012floyd} are hence a natural candidate for describing input-output specifications for quantum programs. 
However, contrary to the classical setting where pre/post conditions have a lot of flexibility in description, the conventional pre/post conditions in quantum Hoare triples are described by quantum predicates which are Hermitian matrices of exponential dimensions in terms of the system size. 
The exponential dimension of quantum predicates incurs both the scalability issue and technical inconvenience in automating the reasoning directly based on quantum Hoare triples. 

Moreover, for \sqirs programs, one needs to express the specifications for a family of programs for different sizes, which is like classical program synthesis with a varying-length array of variables. 
However, existing Hoare triples can only be used to provide specifications for quantum systems of a fixed dimension.

To that end, we develop the so-called \emph{hypothesis-amplitude} specifications for quantum circuit families, where the hypothesis component (denoted $h$) of the specification describes a certain subset of input states $x$ in the computational basis, and the amplitude component (denoted $\alpha$) describes the output state of the program on the given input $x$.  
Both $h, \alpha$ are functions of the index $n$ so that they can describe  a family of quantum circuits.

\begin{definition}
\label{def:judge}
A \emph{hypothesis-amplitude triple} contains $h, S$ and $\alpha$.
Here $h$ is a set of tuples $(n, x, y)\in\mathbb{N}^3$ (we abuse the notation $h$ to also represent its indicator function of type $\mathbb{N}^3\rightarrow\mathbb{B})$ that specifies the interested entries of $S$' semantics, $S$ is a quantum circuit family parameterized with a natural number $n$, and $\alpha(n, x, y)$ is a complex function with natural number inputs.

A hypothesis-amplitude triple is a valid judgement, denoted  $\interpret{h}{S}{\alpha}$, when  
\begin{align}
    \interpret{h}{S}{\alpha} \iff \forall (n, x, y)\in h,~ \bra{y}\sem{\ssem{S}(n)}\ket{x}=\alpha(n, x, y).
\end{align}
\end{definition}

Following the above definition, the hypothesis $h$ is like classical pre-conditions and specifies the set of inputs where the post-conditions are provided. 
For any such input $x$, the output state of the program $S(n)$ is given by $\sem{\ssem{S}(n)}\ket{x}= \sum_y \alpha(n,x,y) \ket{y}$, which explains why $\alpha$ is called the amplitude. We do not always need to specify the program's semantics for all inputs, so we use set $h$ to filter those unnecessary information.
By the linearity of unitary, the input-output specification on a set of inputs in the computational basis can be extended to a specification in the linear space spanned by the given input set.

Compared with quantum Hoare triples, our hypothesis-amplitude specification provides a more flexible and arguably more intuitive way to formalize the desired properties on the target functions. 
For instance, for state preparation, our specification is almost straightforward to use and avoids the extra efforts of converting specifications into exponential-size quantum predicates. 
\footnote{Quantum Hoare triples, however, need a post-predicate where the target state spans the subspace with eigenvalue 1, and the rest space has eigenvalue 0.
These complication comes from the generality of quantum Hoare triple.}
Moreover, for all unitary programs, our hypothesis-amplitude specification could provide the same expressive power as general quantum Hoare triples at the cost of using potentially complicated $h, \alpha$. 
Nevertheless, efficiently encoding into SMT instances are only known in restricted cases of $h, \alpha$ as discussed in Section~\ref{sec:encoding}. 

\subsection{From \name-spec to Hypothesis-amplitude Specification}
\label{sec:spec2ha}
Given a \name-spec specification $I\mapsto O$, \name compiles it to the hypothesis-amplitude specification in two steps: (1) generates the corresponding hypothesis $h$ and a variable map $\Pi$ based on $I$, where $\Pi$ maps variables claimed in $I$ to the qubits; (2) generates the corresponding amplitude function $\alpha$ based on $O$ and $\Pi$.
\paragraph{Generate $\Pi$ and $h$ from $I$} 
\name first calculates the total number of input qubits by adding up the length specifications of constants and variables in the input. This number depends on the parameter $n$. Then, \name{} assigns indexes from low to high for each variable and constant in $I$, according to the order they appear in $I$. The assignment of the variables is included in the variable map $\Pi$. For each constant number $c$ represented by qubits $q_a\sim q_b$, \name adds expression $x[b:a]=c$ into the hypothesis $h$. For example, consider the specification for the quantum adder circuit $I:\ket{0}\ket{A[n]}\ket{B[n]}\ket{0_n}$, the variable map $\Pi$ and the hypothesis $h$ generated by \name are: 
\begin{align}
    \Pi= \{A \mapsto q_1\sim q_n, B \mapsto q_{n+1}\sim q_{2n}\},\qquad h = \{x[0]=0, x[3n:2n+1]=0\}
    \label{equ:pi-ex}
\end{align}
\paragraph{Generate $\alpha$ from $O$ and $\Pi$}
\name will then compute the value $\alpha(n, x, y)$ from the output $O$ and the variable map $\Pi$. \name first calculates the number of qubits. Then, it calculates the normalization factor $k$ which is the total number of qubits in the summed variable plus the number of additions.  Then, \name evaluates the output into a basis state $eval_{\Pi, x, y}(O)$ by replacing input variables with slices of $x$ according to $\Pi$ and summed variables with corresponding slices of $y$ together with a phase factor $e^{i \phi(O, \Pi, x, y)}$. Finally, the value of $\alpha(n, x, y)$ will be $1/\sqrt{2^k} e^{i \phi(O, \Pi, x, y)} \cdot \delta(eval_{\Pi, x, y}(O) = y)$. If the output contains additions, there will be a basis state and a phase factor for each term and the $\alpha$ value will be the sum of their $\alpha$. For example, the $\alpha$'s of GHZ and QFT in Equation~\ref{equ:qftspec} are
\begin{align}
    \alpha_{GHZ} (n, x, y) = \frac{1}{\sqrt{2}}\bvalue{y=0\lor y=2^{n+1}-1}, \qquad \alpha_{QFT} (n, x, y) = \frac{1}{\sqrt{2^n}} e^{2\pi i x \cdot y / 2^n}
\end{align}

\section{INDUCTIVE \sqir AND ITS LOGIC}
\label{sec:formulation}

\name{}'s goal is to synthesize programs with inductive structures.
To that end, we extend the existing intermediate representation \sqir \cite{hietala2021voqc} into a language called Inductive SQIR (\sqirs) that defines a family of quantum circuits inductively in Section~\ref{sec:inductive_sqir}. 
In Section~\ref{sec:logic}, we also develop a logic for reasoning about an \sqirs program with respect to an $h-\alpha$ specification.  

\subsection{Inductive \sqir (\sqirs)} \label{sec:inductive_sqir}
We extend \sqir with an inductive structure, similar to \coqe{fixpoint} in Coq,  to equip the language with the ability to describe a family of quantum circuits for general input $n$. 

\begin{definition}[Inductive \sqir - Syntax]
\label{def:syntax}
An \sqirs program is defined inductively by:

\begin{align*}
    S ::= ~~& \const{P} \mid \seq{S_1}{S_2} \mid \relabel{\pi}{S} \mid \fix_{k}~\pi ~P_0 ~P_1 ~\cdots ~P_{k-1} ~S_L ~S_R.
\end{align*}
Here, $P, P_0, ..., P_{k-1}$ are \sqir programs, $\pi$ is a series of injective natural number mappings.

\end{definition}
At a high level,  any \sqirs program is a succinct way to describe a series of \sqir programs indexed by an integer (or input size) $n=0, 1, 2, \cdots$. 
Intuitively, $\const{P}$ represents a repeating series of \sqir programs where every entry in the series is the same \sqir program $P$.
$\seq{S_1}{S_2}$ concatenates two series $S_1$ and $S_2$ by concatenating \sqir programs of each entry. We also use $S_1; S_2$ and $\seq{S_1}{S_2}$ interchangeably for notation convenience. $\relabel{\pi}{S}$ permutes the qubit labels for the $n$th entry with permutation $\pi(n)$. 

$\fix_{k}$ is the new \emph{inductive} structure introduced to \sqirs. 
Specifically, $\fix_{k}$ constructs a series of \sqir programs by recursion, with $k$ base cases $P_0,...,P_{k-1}$, and the recursive call for the $n$-th entry is sandwiched by the $n$-th entries of \sqirs programs $S_L$ and $S_R$.
The choice of $\fix_{k}$ is inspired by commonly seen quantum programs and serves as a good syntax guide for synthesis purposes for all the case studies in this paper.

We formulate the \emph{semantics} of \sqirs programs as functions from a natural number to   a \sqir program, i.e., $\mathbb{N} \rightarrow \sqir$.

\begin{definition}[\sqirs - Semantics]
An \sqirs program represents a series of unitary \sqir programs $\{P_0, P_1, ...\}.$ We define the IR semantics $\ssem{S}$ of \sqirs program inductively by:
\begin{align*}
    &\ssem{\textbf{const}\ P}(n) = P; \qquad
    \ssem{\textbf{seq}\ S1\ S2}(n) = \ssem{S1}(n) ; \ssem{S2}(n); \\
    &\ssem{\textbf{relabel}\ \pi\ S}(n) = \texttt{map\_qb}(\pi(n), \ssem{S}(n)); \\
    &\ssem{\textbf{fix}_k}(n) = \begin{cases} P_n, & n<k \\ \ssem{S_L}(n) ; \texttt{map\_qb}(\pi(n),\ssem{\textbf{fix}_k}(n-1));\ssem{S_R}(n); & \text{else} \end{cases}
\end{align*}
\end{definition}
Here \texttt{;} is the sequential construct in \sqir, \texttt{map\_qb} is a function relabeling the indices of qubits in a \sqir program according to injective function $\pi(n)$, and $\textbf{fix}_k$ is an abbreviation of $\textbf{fix}_k~\pi\ P_0\ ... \ P_{k-1}\ S_L\ S_R$. 
We use a slightly changed denotational semantics of any \sqir program $P$, denoted $\sem{P}$, which is an \textbf{infinite-dimensional} matrix (i.e., $\mathbb{N}^2\rightarrow\mathbb{C}$), that returns entries of $P$'s original semantics within $P$'s dimension and 0 otherwise.

\begin{example}
As an example, recall the GHZ program written in \sqirs syntax:
\begin{align*}
    \textbf{fix}_1~Id~(\const{H 0})~(\const{ID})~(\relabel{\pi}{(\const{CNOT\ 0\ 1})})~
\end{align*}
where $Id$ is an identity map, $(\const{H 0})$ is a Hadamard gate on qubit $q_0$ which is the $P_0$, $(\const{ID})$ is an identity unitary (served as $S_L$), and $\pi(n)=\lambda x. x+n-1$ for $n\geq 1$. $\pi(n)$ maps the CNOT gate (served as $S_R$) on qubits $q_0,q_1$ to a CNOT gate on qubit $q_{n-1},q_n$. 

For notation convenience, when relabeling a \sqir program with a map $\pi$ (i.e.  $\relabel{\pi}{\const{P}}$), we usually omit the $\mathbf{relabel}$ key word. Thus, $S_R$ can also be denoted as $\const{CNOT n-1 n}$. 

\end{example}

We also develop the following syntax for general permutation $\pi$ used in \sqirs programs that is also part of the candidate program search space. 

\begin{definition}[Permutation Syntax] \label{def:permutation}
\begin{align*}
    \pi ::= ~~& Id \mid \permute{e_1}{e_2} \mid \shift{e_1}{e_2}{m} \mid \pi_1\cdot \pi_2 \\
    e ::= ~~& n \mid m \mid e_1 + e_2 \mid e_1-e_2 \mid e_1*e_2 ~\qquad m\in \nat
\end{align*}
$Id$ is an identity permutation; $\permute{e_1}{e_2}$ swaps the mapping $e_1$ and $e_2$. ${\shift{e_1}{e_2}{m}}$ maps any $x\in[e_1,e_2)$ to $[ (x-e_1+m)\text{ mod } (e_2-e_1)] + e_1)$.

\end{definition}
For example, permutation $\permute{1}{n}(n)$ maps 1 to $n$ and maps $n$ to 1. Permutation $({\shift{2}{5}{1}})(n)$ maps 4 to 2, 2 to 3, and 3 to 4.

\subsection{Unitary \sqirs Logic}
\label{sec:logic}

We develop \textit{unitary \sqirs logic} shown in Fig~\ref{fig:ulogic} to reason about \sqirs program's semantics with respect to the hypothesis-amplitude specification $(h,\alpha)$ with helper functions 
in Def~\ref{def:calculation}. The soundness is formally proven in Theorem~\ref{thm:soundness}, whose proof is postponed to Appendix A.1 in the supplementary material.

One can view $(h, \alpha)$ as a series of incomplete matrices: only those entries in the set $h$ are known, whose values can be looked up in $\alpha$. This gives the intuition behind our rules. The \textsc{Weaken} rule states that any subset of $h$ can also be observed by $\alpha$. The \textsc{Const} rule lifts \sqir semantics to \sqirs semantics. The \textsc{Replace} rule states that if the observed entries are the same for two amplitude functions, then one can substitute the other one with hypothesis $h$. The \textsc{Relabel} rule relabels the entries of matrices for both $h$ and $\alpha$. The \textsc{Seq} rule calculates matrix multiplication for each term in the series. 

The \textsc{Fix} rule checks observed entries for terms with index $i<k$ (base cases) and computes matrix multiplication for $i\geq k$ (inductive cases).

\begin{definition}
\label{def:calculation}
For a hypothesis set $h$ and an amplitude function $\alpha$, 

the \textbf{relabeling function} and the \textbf{predecessor functions} of $h$ and $\alpha$ are defined by
\begin{gather*}
\pi\circ(h, \alpha):=(\pi\circ h, \pi\circ \alpha) \qquad \text{pred }(h, \alpha):=(\text{pred }h, \text{pred }\alpha) \\
        \pi \circ h :=\{ (n,\pi(n,x),\pi(n,y))\mid(n,x,y)\in h\}\quad
    (\pi \circ \alpha)(n,x,y) := \alpha(n,\pi(n,x),\pi(n,y)) \\
    (\text{pred }\alpha)(n, x, y):=\alpha(n-1, x, y)\qquad
    \text{pred }h:=\{(n-1, x, y) \mid (n,x,y)\in h\}
\end{gather*}
The entries of $h$ and $\alpha$ are relabeled according to a series of injective function $\pi$. Predecessor functions move the series by one index, and are used for recursive calls in fixed-point programs.

The \textbf{composition function} of $h$ and $\alpha$ are defined by:
\begin{align*}
    \text{comp}(h_1,\alpha_1,h_2,\alpha_2)&:=\Big\{(n, x, y):\forall z, \left((n, x, z)\in h_1\wedge (n, z, y)\in h_2\right)\vee \nonumber \\
    &((n, x, z)\in h_1\wedge\alpha_1(n, x, z)=0) \vee\left((n, z, y)\in h_2\wedge\alpha_2(n, z, y)=0\right)\Big\},\\
    (\alpha_1 * \alpha_2)(n, x, y)&:=\sum_{z\in\mathbb{N}} \alpha_1(n, x, z)\alpha_2(n, z, y),\\
    (h_1, \alpha_1) \otimes (h_2,\alpha_2)&:=(\text{comp}(h_1,\alpha_1,h_2,\alpha_2), \alpha_1 * \alpha_2).
\end{align*}

We also define several restricted \textbf{equivalence relations}:
\begin{align*}
  &h_1\equiv_n h_2 \Leftrightarrow (\forall x, \forall y, (n, x, y)\in h_1\leftrightarrow (n,x,y)\in h_2) \\
  &\alpha_1 \equiv^{h}_n \alpha_2 \Leftrightarrow \forall x, \forall y, (n, x, y) \in h\rightarrow \alpha_1(n, x, y) = \alpha_2(n, x, y) \\
 & \alpha_1 \equiv^{h} \alpha_2\Leftrightarrow\forall n, \alpha_1 \equiv^{h}_n \alpha_2, \qquad(h_1, \alpha_1)\equiv_n (h_2, \alpha_2) \Leftrightarrow h_1\equiv_n h_2 ~\wedge~ \alpha_1\equiv^{h_1}_n \alpha_2
\end{align*}

\end{definition}

\begin{figure}
\centering\small
    \begin{minipage}{0.4\linewidth}
    \centering
    \begin{align*}
        \inferrule*[Right=\textsc{Weaken}]{\interpret{h}{S}{\alpha}, \quad h'\subseteq h}{\interpret{h'}{S}{\alpha}}
    \end{align*}
    \begin{align*}
        \inferrule*[Right={Replace}]{\interpret{h}{S}{\alpha}, \quad  \alpha\equiv^h\alpha'}{\interpret{h}{S}{\alpha'}}
    \end{align*}
    \begin{align*}
        \inferrule*[Right=Seq]{
    \substack{\interpret{h_i}{S_i}{\alpha_i} \quad \forall i=1, 2 \\
    (h, \alpha)=(h_1, \alpha_1)\otimes(h_2, \alpha_2)}}{\interpret{h}{\textbf{seq }S_1~S_2}{\alpha}}
    \end{align*}
    \end{minipage}
    \qquad
    \begin{minipage}{0.4\linewidth}
    \centering
    \begin{align*}
        \inferrule*[Right=Const]{ \alpha\equiv^{{\mathbb N}^3}\lambda n.\sem{P}}{\mathbb{N}^3\triangleright (\textbf{const } P)\leftrightarrow\alpha}
    \end{align*}
    \begin{align*}
        \inferrule*[Right=Relabel]{
    {\interpret{h}{S}{\alpha}\quad \alpha'\equiv^{\pi \circ h}\pi \circ \alpha}
    }{\interpret{\pi \circ h}{\textbf{relabel }\pi~S}{\alpha'}}
    \end{align*}
    \begin{align*}
        \inferrule*[Right=Fix]{
    \substack{
    \interpret{h_L}{S_L}{\alpha_L}, \quad \interpret{h_R}{S_R}{\alpha_R} \\
    \forall i<k, \quad \interpret{h_i}{\textbf{const}~P_i}{\alpha_i}, \quad (h, \alpha)\equiv_i (h_i, \alpha_i) \\
    \forall i\geq k, \quad (h, \alpha)\equiv_i (h_L, \alpha_L)\otimes (\pi\circ\text{pred }(h, \alpha)) \otimes (h_R, \alpha_R)
    }
    }
    {\interpret{h}{\textbf{fix}_k~P_0~\cdots~P_{k-1}~S_L~S_R}{\alpha}}
    \end{align*}
    \end{minipage}
    \caption{The unitary \sqirs logic.}
    \label{fig:ulogic}
\end{figure}

\begin{theorem}
\label{thm:soundness}
The rules of unitary \sqirs logic in Fig~\ref{fig:ulogic} are sound.
\end{theorem}

\section{EFFICIENT ENCODING BY PARAMETERIZED PATH-SUM AMPLITUDE}
\label{sec:encoding}

In this section we explain how to encode the \logicName triple introduced In Definition~\ref{def:judge} and the functions and relations in Definition~\ref{def:calculation} into the SMT instance.\par
\subsection{Encoding The Hypothesis-amplitude Triple}
The \logicName triple in Definition~\ref{def:judge} includes a hypothesis set $h$ and an amplitude function $\alpha$, which will be encoded separately. 
\paragraph{Encoding the hypothesis set}
The hypothesis set $h$ refers to a set of natural numbers.
%and its encoding is intuitive:
Intuitively, we encode the hypothesis $h$ with a Boolean expression $B$ constructed by $n,x,y$, whose value is true if and only if $(n,x,y)$ is inside the hypothesis set.  Namely, 
% We encode $h$ in the form
%  \begin{align}
%     h:\{(n,x,y) \mid B(n,x,y) \}
% \end{align}
% where $B(n,x,y)$ is a \textit{Boolean} expression constructed by the variables $n,x,y$. The hypothesis in this form suggests we only focus on the entries $(n,x,y)$ that makes the Boolean expression $B(n,x,y)$ true. By encoding $h$ in this form, the term $(n,x,y)\in h$ can be encoded as
\begin{align*}
    (n,x,y)\in h \Longleftrightarrow B(n,x,y)=\textbf{True}.
\end{align*}

\paragraph{Encoding the amplitude function}
 \label{sec:encode-alpha}
Encoding the complex function $\alpha$ is challenging since there is currently no automated program verification tool that supports complex numbers.
We solve this by restricting the function $\alpha$ in a limited form that can be encoded into SMT instances. This is a trade-off between the expressiveness of our specification and the feasibility of automated verification. \par

\paragraph{Parameterized path-sum amplitude (\amplfunc) function}
We restrict an amplitude function $\alpha$ to be a \textit{Parameterized Path-Sum Amplitude} function:
%For the amplitude function $\alpha$, we restrict $\alpha$ to be a \emph{parametrized path-sums amplitude} function (\amplfunc function) defined below
 \begin{definition} \label{def:ppsa}
 A Parameterized Path-Sum Amplitude (\amplfunc) function $\alpha_p:\nat^3 \rightarrow \mathbb{C}$ is defined as 
  \begin{align*}
    \alpha_p(n,x,y)&:= \frac{1}{\sqrt{\beta(n)}} \sum_{i=0}^m \dvexp{\bvalue{B_i(n,x,y)}}{V_i(n,x,y)}
    %\\S(n)&:\ket{x}\mapsto \frac{1}{\sqrt{2}^{\gamma(n)}} \sum_{y\in \mathbb{N}} \sum_{i=0}^m \delta_i(n,x,y) \cdot e^{2\pi i P_i(n,x,y)} \ket{y}
\end{align*}
\begin{itemize}
    \item $\beta$ is a natural number expression of $n$ and it decides the magnitude of all paths.
    \item $m\in \nat$ is a constant number. $\{B_i\}_m$ is a group of boolean expressions constructed by $(n,x,y)$ and satisfies $\forall n~x~y\in \nat, \sum_{i=0}^m \bvalue{B_i(n,x,y)} \le 1$, where $\bvalue{B}:\text{Bool}\rightarrow \{0,1\}$ is a function that returns 1 if $B$ is True and returns 0 otherwise. 
    \item $\{V_i\}_m$ is a group of natural number expressions constructed by $(n,x,y)$.
    \item  For $x\in \mathbb{N}$, suppose $x$'s binary representation is $x_q...x_1x_0$ where $x_i\in \{0,1\}$, we use $d(x)$ to denote the fractional binary notation of $x$.
    \begin{align*}
    d(x) = [0.x_0x_1...x_q]_2 = \sum_{i=0}^q x_i \cdot 2^{-(i+1)}.
\end{align*}  
\end{itemize}
 \end{definition}
% Since we focus on states in which all possible paths' amplitudes have the same magnitude, $\beta$ only depends on $n$ and does not depend on $x,y$. Since function $\alpha$ has non-zero value \textit{iff} $B(n,x,y)$ is true, the Boolean expression $B$ decides exactly the path set $T$ as $\{y \mid B(n,x,y)\}$. 
% The phase of each path $e^{-2\pi i \cdot d(V(n,x,y))}$ is determined by the expression $V$. 

%For the expression $V$, it decides the phase of each path's amplitude.
 
 Fig.~\ref{fig:halpha} shows the syntax we allow to construct the expressions $\beta, B, V$ in a PPSA function.  "$V_1{~\textbf{rop}~}V_2$" is a set of common relational operators bewteen $V_1,V_2$(e.g. $=~\neq~>~<~\ge~\le$). "$\text{\textbf{uop}~}V$" is a set of common unary operator on $V$(i.e. -~!~\&~|~ $\oplus$). "$V_1\text{~\textbf{bop}~}V_2$" is a set of binary operators between $V_1$ and $V_2$, including arithmetic operators (i.e. + - * $\backslash$ \% ) and bit-wise operators (i.e. $\&~|~ \oplus$ << >>). All these operators have the same meaning as they have in C language. $V_1[V_2]$ means the $V_2$-th bit of $V_1$'s binary representation (in the order from low to high). $v'[V_1:V_2]$ is the natural number represented by the binary representation truncated from high bit $V'_{V_1}$ to low bit $V'_{V_2}$. For example, let $V_1=(6)_{10}=(110)_2$, we have
$V_1[0]=0, V_1[2:1]=(11)_2=3$.
Z3 SMT solver supports all these syntaxes. \par

\begin{figure}
    \centering\small
    \begin{tabular}{cc}
     $n,x,y: \text{Variables}\in \mathbb{N}$ & $k:\text{Fixed number}\in \mathbb{Z}$  
    \end{tabular}
\begin{align*}
    %\text{Hypothesis }h&::=\quad \{(n,x,y)|B(n,x,y)\}\\
    % \text{Amplitude Function }\alpha &::=\quad \lnxy ~ \frac{1}{\sqrt{\beta(n)}}\cdot \dvexp{\bvalue{B(n,x,y)}}{V(n,x,y)}\\
    \text{Boolean Expression }B & ::=\quad \text{\textbf{True}} \mid \text{\textbf{False}} \mid B_1 \land B_2 \mid B_1 \lor B_2 \mid \neg B' \mid V_1{~\textbf{rop}~}V_2 \\
    \text{Binary-}\mathbb{N}\text~V & ::=\quad x \mid y \mid n \mid k \mid \delta(B) \mid V_1[V_2] \mid V'[V_1:V_2] \mid \text{\textbf{uop}~}V' \mid V_1\text{~\textbf{bop}~}V_2 \\  % \mid \mask(V_1,V_2,V_3) \mid  \ext(1,v) 
    \text{Magnitude }\beta &::= \quad k \mid n \mid \beta_1\text{~\textbf{bop}~}\beta_2 \mid 2^n
\end{align*}   
\vspace{-10pt}
\caption{Syntax of the
PPSA function. 
%$\mask(V_1,V_2,V_3)$ means to set from the $V_2$-th to the $V_3$-th bits in $v_1$ to 0 (e.g. \mask($(11111)_2$, 0,1)=$(11100)_2$).
%$\ext(1,v)$ constructs a binary number by repeating $v$ copies of bit 1 and its value equals to $2^v-1$. 
}
\vspace{-10pt}
\label{fig:halpha}
\end{figure}
By restricting amplitude function $\alpha$ to a \amplfunc function, we disassembled the complex number function $\alpha$ into the combination of several integer or boolean expressions, which allows us to represent $\alpha$ with a set of SMT expressions that enable us to encode the calculation in Definition~\ref{def:calculation} into the SMT solver. This will be discussed in Section~\ref{sec:encode-cal}.

Our design for the \amplfunc function is inspired by Feynman’s \textit{sum-over-path} formalism described in Section~\ref{sec:path-sum} 
which has inspired many quantum state representations. 
However, all of these representations can only express constant size unitary operators and fail to work for any input size. PPSA inherits the expressibility of the existing sum-over-path representations, which can express most famous quantum algorithms (e.g.,~\cite{amy2018towards,qbricks2020deductive}), and works for a general input size. Hence, we believe the restriction to PPSA is mild and serves as a good balance between expressiveness and feasibility. 
Some common unitary operators that can be represented by h-$\alpha$ triple while restricting the amplitude function to \amplfunc are listed in Table~\ref{tab:example}. More examples are provided in Section~\ref{sec:case}.
 
\begin{table}[htbp]
 \caption{Examples of Unitary Operators represented by the H-$\alpha$ Specification.} \label{tab:example}
\small
\begin{tabular}{|c|c|c|} %\label{tab:expressiveness}
 \hline
     \textbf{Name} & \textbf{Unitary Operator} & \textbf{H-$\alpha$ Specification} \\
    \hline
     $\text{Uniform}_{n+1}$ & $\ket{0}^{n+1} \mapsto \frac{1}{\sqrt{2^{n+1}}} \sum_{0\le y < 2^{n+1}} \ket{y}$ & \makecell{$h= \{(n,x,y) | x=0\}$\\ $\alpha(n,x,y)= \frac{1}{\sqrt{2^{n+1}}} \bvalue{y<2^{n+1}}$}\\
     \hline
     $\text{Toffoli}_{n+1}$ & \makecell{$\ket{q_0q_1\cdots q_n} \mapsto$\\$\ket{q_0q_1\cdots (q_n \oplus \prod_{i=0}^{n-1}q_i))}$} & \makecell{$h= \{(n,x,y) | x<2^{n+1} \land y<2^{n+1}\}$\\$\alpha(n,x,y)= \bvalue{x[n-1:0]=y[n-1:0] \land$ \\ $y[n]=x[n]\oplus(\&x[n-1:0])}$} \\
     \hline
    $\text{QFT}_{n+1}$ & $  \ket{x} \mapsto \frac{1}{\sqrt{2^n}} \sum_{y=0}^{2^n-1} e^{\frac{2\pi i\cdot xy}{2^n}} \ket{y}$ & \makecell{$h= \{x<2^{n+1}\land y<2^{n+1} \}$\\
    $\alpha(n,x,y)= \frac{1}{\sqrt{2}^{n+1}} \cdot \phase{(x\cdot y)\text{>>}(n+1)}$}\\
    \hline
 \end{tabular}
\end{table}

\subsection{Encoding Reasoning Based on \sqirs Logic}
\label{sec:encode-cal}

To enable SMT-based automation in reasoning, 
one needs to encode the functions and the equivalence relations of $h,\alpha$ defined in Definition~\ref{def:calculation} into SMT instances. 
In the cases of the relabeling functions $(\pi\circ h), (\pi \circ \alpha)$, the predecessor function $\text{pred }(h,\alpha)$ and the composition function $\text{comp}(h_1,\alpha_1,h_2,\alpha_2)$, all used operations are supported by SMT solvers directly and the encoding is trivial. We hence focus our discussison on the non-trivial encoding of the composition function $\alpha_1*\alpha_2$ and the equivalence relations.\par

\paragraph{Encoding the composition function}
Recall the $\alpha_1*\alpha_2$ function from Definition~\ref{def:calculation}:
\begin{align*}
    (\alpha_1*\alpha_2)(n,x,y)=\sum_{z\in \nat} \alpha_1(n, x, z) \alpha_2(n, z, y).
\end{align*}
Since the summation is over $z\in \mathbb{N}$, by definition, the $\alpha_1*\alpha_2$ function is a composition of two infinite-dimension unitaries, and hence cannot be calculated directly. 

All existing symbolic matrix multiplication methods can only deal with a fixed dimension or a fixed number of terms (e.g.,~\cite{amy2018towards}) and hence are not applicable in our case.

Fortunately, we observe that in many cases, non-zero values of the function $\alpha$ are sparse,  

making the composition possible. \emph{In particular, we show the possibility of computing the function $\alpha_1*\alpha_2$ when one of $\alpha_1$ or $\alpha_2$ is sparse}. The sparsity of $\alpha$ is precisely defined as
\begin{definition} \label{def:sparsity}
 We say a function $\alpha:\nat^3\rightarrow\mathbb{C}$ is \textbf{sparse} iff: there exist two functions $\X, \Y: \mathbb{N}^2 \rightarrow \{\mathbb{N}\}$ and for any inputs, the sets returned by $\X, \Y$ always have constant sizes (i.e., independent of inputs $n,x,y$), and further satisfy 
\begin{align*}
    &\forall n~x~y \in \nat,\quad \alpha(n,x,y)\neq 0 \rightarrow x\in \X(n,y) \land y\in \Y(n,x).
\end{align*}
We denote such sparsity by $\sparse{\alpha}{\X}{\Y}$.
\end{definition}

Intuitively, when $\sparse{\alpha}{\X}{\Y}$ holds, for any given $n_0,x_0\in \nat$, 
$\alpha(n_0,x_0,y)$ has non-zero values only on a finite of $y$ points, the set of which is  $\Y(n_0,x_0)$. 
The same intuition holds for $\X$ except for the case when $n_0, y_0$ are given. 
\begin{example}
We show the amplitude function that can represent the \sqirs program $\const{H 0}$ and its sparsity tuple $\X,\Y$ as an example.
\begin{align*}
    &\interpret{\nat^3}{\const{H 0}}{\alpha_H},\qquad \alpha_H(n,x,y)=\frac{1}{\sqrt{2}}\bvalue{x\backslash 2=y\backslash 2}\cdot e^{2\pi i\cdot \frac{x[0]*y[0]}{2}}\\ &\sparse{\alpha_H}{\X}{\Y}, \qquad \X(n,y)=\{y, y\oplus 1\},\qquad \Y(n,x)=\{x, x\oplus 1\}.
\end{align*}
The operation $\oplus$ is a bit-wise operation and the expression $x\oplus 1$ flips the 0th bit of $x$ (e.g. $(101)_{2} \oplus 1 = (100)_{2}=4$). The expression $\bvalue{x\backslash 2=y\backslash 2}$ in $\alpha_H$ indicates that
\begin{align*}
    \forall n~x~y \in \nat,\quad\alpha_H(n,x,y)\neq 0 \rightarrow x\in \X(n,y) \land y\in \Y(n,x)
\end{align*}
Intuitively, $\X,\Y$ are constructed in this way since $\const{H 0}$ only modifies the 0th qubit.
\end{example}

Now we explain how to encode $\alpha$ function $\alpha_1*\alpha_2$ when one of $\alpha_1, \alpha_2$ is sparse. Suppose $\alpha_2$ is sparse and we have $\sparse{\alpha_2}{\X}{\Y}$, we know that $\alpha_2(n,z,y)\neq 0$ only when $z\in \X(n,y)$. So $\alpha_1*\alpha_2$ can be calculated by
\begin{align*}
    (\alpha_1*\alpha_2)(n,x,y)=\sum_{z\in \nat} \alpha_1(n,x,z)\alpha_2(n,z,y) = \sum_{z \in \X(n,y)} \alpha_1(n,x,z) \alpha_2(n,z,y).
\end{align*}
The summation on the right hand has only a fixed number of terms by sparsity which allows encoding into SMT instances. Similarly, when $\alpha_1$ is sparse and $\sparse{\alpha_1}{\X}{\Y}$, we have
\begin{align*}
    (\alpha_1*\alpha_2)(n,x,y)=\sum_{z\in \nat} \alpha_1(n,x,z)\alpha_2(n,z,y) = \sum_{z \in \Y(n,x)} \alpha_1(n,x,z) \alpha_2(n,z,y).
\end{align*}

Moreover, sparsity of $\alpha$ can be established in many cases. (Proof in Appendix A.1).

\begin{theorem}[$\alpha$-sparsity] Suppose $\alpha, \alpha_1, \alpha_2$ are amplitude functions:
\label{theo:sqir}
\begin{itemize}
    \item Let $P$ be a unitary \sqir program and $\interpret{\nat^3}{\const{P}}{\alpha}$, then $\alpha$ is sparse.
    \item If $\alpha$ is sparse and $\pi$ is a series of injective natural number mappings, then $\pi\circ \alpha$ is sparse.
    \item If both $\alpha_1, \alpha_2$ are sparse, so is  $\alpha_1*\alpha_2$.
\end{itemize}
\end{theorem}

The above theorem shows that the $\alpha$ functions for all SQIR programs, and for relabeling a SQIR program or composing two SQIR programs are sparse. 
So non-sparse $\alpha$s only appear in the fixpoint syntax. 
The candidate program from our searcher has at most one fixpoint due to the challenge discussed in Section~\ref{sec:future}. So when composing two amplitude functions $\alpha_1,\alpha_2$, there is always at least one sparse function and our composition strategy can work.\par

\paragraph{Encoding the equivalence relations}
Given a hypothesis $h$ and two complex functions $\alpha,\alpha'$, 
suppose the functions $\alpha,\alpha'$ are in the form:
\begin{align*}
    \alpha(n,x,y)=\frac{1}{\sqrt{\beta(n)}}\sum_{i=0}^{m}\dvexp{\bvalue{B_i(n,x,y)}}{V_i(n,x,y)}\\
    \alpha'(n,x,y)=\frac{1}{\sqrt{\beta'(n)}}\sum_{i=0}^{m'}\dvexp{\bvalue{B'_i(n,x,y)}}{V'_i(n,x,y)}\\
\end{align*}
\normalsize

\name verifier checks the equivalence $\ufeq{h}{\alpha}{\alpha'}$ by the checking following SMT instance and rejects the equivalence when the SMT solver gives a negative result.

\begin{align*}
    \forall n~x~y\in \nat, ~~ h(n, x, y)\rightarrow & \beta(n)=\beta'(n) \land \sum_{i=0}^{m}B_i(n,x,y) = \sum_{i=0}^{m'}B'_i(n,x,y) \\
    &\land \sum_{i=0}^{m}B_i(n,x,y)*V_i(n,x,y)=\sum_{i=0}^{m'}B'_i(n,x,y)*V'_i(n,x,y).
\end{align*}

\section{EXPERIMENTAL CASE STUDIES} \label{sec:case}

We demonstrate six additional case studies and provide the output Qiskit programs compiled from synthesized \sqirs programs for better illustration.
Then in Section~\ref{sec:performance}, we compare the performance of \name against the previous quantum circuit synthesis frameworks, QFAST~\cite{younis2021qfast} and Qsyn~\cite{kang2023modular}.

\subsection{Quantum Adder}
\label{sec:adder}
\paragraph{Motivation and Background}
Quantum circuits for arithmetic operations are required for quantum algorithms. One important example is the adder circuit.  \citet{feynman1985quantum} first proposes the quantum \emph{full adder} circuit to implement  $\ket{0}\ket{A}\ket{B}\ket{0}^{\otimes n} \rightarrow \ket{c_0}\ket{A}\ket{B}\ket{A+B}$ where $A,B$ are $n$-bit natural number. The first $\ket{0}$ is the carry bit and it is changed to carry value $\ket{c_0}$ after the addition. 
This design unfortunately needs $n$ more qubits to store the sum of $A+B$. 
To reduce the qubit usage,  \citet{cuccaro2004new} proposed a new \emph{ripple-carry adder} that 

uses $n$ fewer qubits than the full adder design.  When given different specifications, \name can synthesize both adder circuits.
\paragraph{Full Adder Synthesis}
We let \name synthesize a program $S_a$ that
$S_a(n)$ provides a $n$-qubit full quantum adder (i.e. $S_a(0)$ is an identity unitary) with the specification in Equation~\ref{equ:adderspec}. \name generates a program $S_a$ as shown in Fig~\ref{fig:fulladder}(a)(c) (i.e. Qiskit function \texttt{full\_adder}). When $n=0$, $S_a$ does nothing since the circuit only contains the carry bit. When $n\ge 1$, $S_a$ first call $S_a(n-1)$ recursively to get a $n-1$-bit full adder to calculate $\ket{A_{n-1}+B_{n-1}}$ and the carry bit is stored in qubit 0. Then $S_a$ uses the one-bit adder circuit $S_R$ to sum the highest bit in $A_n$ and $B_n$.

\begin{figure}
 \begin{minipage}[p]{0.42\textwidth}
 \begin{lstlisting}[language=Python]
def full_adder(N):
    circuit = QuantumCircuit(3N+1)
    def Sa(circ, n):
        if n==0: 
            return
        else:
            Sa(circ,n-1)
            @\textcolor{blue}{\# See the circuit SR below}@
            circ.ccx(n, N+n, 2N+n)
            circ.cx(n, N+n)
            circ.ccx(N+n, 0, 2N+n)
            circ.cx(N+n, 0)
            circ.cx(n, N+n)
            circ.swap(0, 2N+n)
    Sa(circuit,N)
    return circuit
\end{lstlisting}
    \caption*{(a)}
\end{minipage}~
\hspace{1.3em}~
\begin{minipage}[p]{0.45\textwidth}
\begin{lstlisting}[language=Python]
def Cuccaro_adder(N):
    circuit = QuantumCircuit(2*N+1)
    def S(circ, n):
        if n==0: 
            return
        else:
            @\keynum{\# MAJ}@
            circ.cx(N-n+1,2*N-n+1)
            circ.cx(N-n+1,N-n)
            circ.ccx(N-n,2*N-n+1,N-n+1)
            S(circ,n-1)
            @\keynum{\# UMA}@
            circ.ccx(N-n,2*N-n+1,N-n+1)
            circ.cx(N-n+1,N-n)
            circ.cx(N-n,2*N-n+1)
    S(circuit,N)
    return circuit
\end{lstlisting}
    \caption*{(b)}
\end{minipage}\\
    \begin{minipage}[p]{0.3\textwidth}
    \centering
    $S_R$\\    
    \scalebox{0.75}{

\Qcircuit @C=1.0em @R=1.0em {
    \lstick{q_{n}} & \ctrl{3} & \ctrl{1} & \qw & \qw & \ctrl{1} & \qw & \\
    \lstick{q_{N+n}} & \ctrl{2} & \targ & \ctrl{2} & \ctrl{1} & \targ & \qw & \\
    \lstick{q_0} & \qw & \qw & \ctrl{1} &\targ & \qswap & \qw &\\
    \lstick{q_{2N+n}} &\targ& \qw & \targ & \qw & \qswap \qwx& \qw & 
}
}
\caption*{(c)}
    \end{minipage}~
\hspace{0.6em}~
\begin{minipage}[m]{0.71\textwidth}
\scalebox{0.75}{
    \Qcircuit @C=1em @R=0.5em {
      \lstick{0=q_0}  & \multigate{2}{MAJ} & \qw & \qw & \qw & \qw &\qw & \qw & \qw&\qw  & \multigate{2}{UMA}& \qw & \rstick{0}\\
       \lstick{A_1} & \ghost{MAJ} & \qw & \qw & \qw & \qw&\qw & \qw & \qw&\qw  & \ghost{UMA} & \qw & \rstick{A_1}\\
       \lstick{B_1} & \ghost{MAJ} & \multigate{2}{MAJ} & \qw & \qw & \qw&\qw & \qw & \qw& \multigate{2}{UMA} & \ghost{UMA} & \qw & \rstick{S_1}\\ 
       \lstick{A_2} & \qw  & \ghost{MAJ} & \qw & \qw & \qw&\qw & \qw & \qw& \ghost{UMA} & \qw& \qw & \rstick{A_2}\\
         \lstick{B_2}& \qw  & \ghost{MAJ} &\qw & \qw & \qw &\qw & \qw & \qw& \ghost{UMA} & \qw& \qw & \rstick{S_2}\\
         & \vdots & & \ddots & &\textcolor{red}{\text{$S_c(n-1)$}} &  &\begin{sideways}$\ddots$\end{sideways} & &&\vdots\\
         & & & & & & & & & & & & & \\
         \lstick{B_{n-1}}&\qw & \qw & \qw &\qw  & \multigate{2}{MAJ} & \multigate{2}{UMA} &\qw & \qw & \qw &\qw & \qw & \rstick{S_{n-1}}\\
         \lstick{A_n}&\qw & \qw & \qw &\qw  & \ghost{MAJ} & \ghost{UMA} &\qw & \qw & \qw &\qw & \qw & \rstick{A_n}\\
         \lstick{B_n}&\qw & \qw & \qw &\qw  & \ghost{MAJ} & \ghost{UMA} &\qw & \qw & \qw &\qw & \qw & \rstick{S_n}\gategroup{3}{3}{11}{10}{0.7em}{--}\\
   &&&&&&&&&&&&\\ 
    }
}
\caption*{(d)}
\end{minipage}
\vspace{-10pt}
    \caption{(a)(c) Full quantum adder program $S_a$ written in Qiskit. \texttt{circ.ccx(a,b,c)} means appending a Toffoli gate that controlled by qubit $q_a,q_b$ on qubit $q_c$ to the circuit $\texttt{circ}$. The circuit $S_R$ is exactly a one-bit quantum full adder circuit. (b)(d) Cuccaro's quantum ripple-carry adder program written in Qiskit language. 
    }
    \label{fig:fulladder}
\end{figure}
\paragraph{Cuccaro's Adder Synthesis}
To reduce the number of qubits in the circuit, we want to synthesize an in-place adder. The specification is given as:
\begin{align}
    \ket{0}\ket{A[n]}\ket{B[n]}\mapsto \ket{c_0}\ket{A}\ket{A+B}
    \label{equ:radder}
\end{align}

With this specification, \name generates program $S_c$ shown in Fig~\ref{fig:fulladder}(b)(d). We use Cuccaro's MAJ and UMA circuit structures as predefined modules in the synthesis. QSynth flattens these two modules in the compilation process.

\subsection{Quantum Subtractor}
Another important quantum arithmetic operation is the quantum subtractor. A classical $n$-bit subtractor is usually implemented based on the two's complement theory (i.e. $B-A=B+\bar{A}+1$ where  $\bar{A}$ flips each bit in $A$), which is also used by many existing quantum libraries (e.g. QLib~\cite{lin2014qlib},QPanda~\cite{dou2022qpanda}). However, this method requires additional ancilla qubits to build the "+1" operation. To reduce the qubit usage, we let \name synthesize a $n$-bit subtractor using the same number of qubits in the $n$-bit ripple adder.

\paragraph{Synthesis with \name} We let \name synthesize a program with the specification below.

\begin{align*}
    \ket{0}\ket{A[n]}\ket{B[n]}\mapsto \ket{c_0}\ket{A}\ket{B-A}
\end{align*}
This specification is similar to the specification for Cuccaro's Adder (Equation~\ref{equ:radder}) except for changing $B+A$ to $B-A$. With this specification, \name generates the program shown in Fig~\ref{fig:subtractor}. This program equals to the circuit shown in Fig~\ref{fig:subcirc}, indicating the logical expression $B-A=\overline{\bar{B}+A}$, which is different from the traditional subtractor implementation. This subtractor generated by \name uses no ancilla qubits and saves quantum resources.

\begin{figure}
    \centering
    \begin{minipage}[m]{0.45\textwidth}
\begin{lstlisting}[language=python]
def RippleSubtractor(N):
    circuit=QuantumCircuit(2*N+1)
    def S(circ, n):
        if(n==0):
            circ.id(0)
        else:
            circ.x(2*N-n+1)
            circ.cx(N-n+1,2*N-n+1)
            circ.cx(N-n+1,0)
            circ.ccx(0,2*N-n+1,N-n+1)
            S(circ,n-1)
            circ.ccx(0,2*N-n+1,N-n+1)
            circ.cx(N-n+1,0)
            circ.cx(0,2*N-n+1)
            circ.x(2*N-n+1)
    S(circuit,N)
    return circuit
\end{lstlisting}

    \caption{$N$-bit ripple subtractor program written in Qiskit language.}
    \label{fig:subtractor}
    \end{minipage}~\hspace{1em}
    \begin{minipage}[m]{0.45\textwidth}
\begin{lstlisting}[language=python]
def ConditionalAdder(N):
    circ=QuantumCircuit(2*N+2)
    def S(circ, n):
        if(n==0):
            circ.id(0)
        else:
            circ.ccx(0,N-n+2,2*N-n+2)
            circ.ccx(0, N-n+2,N-n+1)
            circ.ccx(N-n+1,2*N-n+2,N-n+2)
            S(circ,n-1)
            circ.ccx(N-n+1,2*N-n+2,N-n+2)
            circ.ccx(0,N-n+2,N-n+1)
            circ.ccx(0,N-n+2,2*N-n+2)
    S(circuit,N)
    return circuit
\end{lstlisting}

    \caption{$N$-bit conditional adder program written in Qiskit language.}
    \label{fig:cadder}
    \end{minipage}
\end{figure}

\subsection{Quantum Conditional Adder}
\paragraph{Motivation and Background}
Quantum Conditional Adder is a necessary arithmetic operation for many known quantum algorithms, including quantum multiplier and \citet{thapliyal2019quantum}'s quantum long division algorithm. A $n$-bit Quantum conditional adder circuit implements the transformation $\ket{ctrl}\ket{0}\ket{A}\ket{B}\mapsto \ket{ctrl}\ket{ctrl*c0}\ket{A}\ket{ctrl*A+ B}$. It sums $A_n$ and $B_n$ when the control qubit $\ket{ctrl}$ is in state $\ket{1}$ and keeps the state unchanged when $\ket{ctrl}$ is in state $\ket{0}$.

One way to construct such a program is by replacing each gate in Cuccaro's adder program (i.e., the program in Fig~\ref{fig:fulladder}(b)) with its conditional version, which is also the circuit generated by Qiskit. However, this method needs four-qubit Toffoli gates, which needs 14 CNOT gates to implement, increasing the total count of CNOT gates in the decomposed circuit. We let QSynth synthesize a target program using only X gate, CNOT gate, and Toffoli gate to find a better solution.

\begin{figure}
    \centering
       \begin{minipage}[m]{0.35\textwidth}
    \centering
    \includegraphics[width=\textwidth]{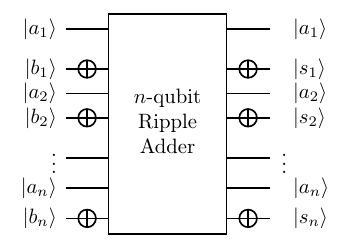}
    \caption{$N$-bit ripple subtractor circuit}
    \label{fig:subcirc}
    \end{minipage}
~\hspace{1em}
    \begin{minipage}[m]{0.42\textwidth}
    \centering
    \small
    \begin{tabular}{c|cc|cc}
    \hline
     \multirow{2}{*}{\textbf{Qubit}} & \multicolumn{2}{c|}{\textbf{QSynth}} & \multicolumn{2}{|c}{\textbf{Qiskit}}\\
    \cline{2-5}
          & $u_3$ & cnot & $u_3$ & cnot\\
        \hline \hline
        4 & 182 & 144 & 244 & 208\\
        \hline
        5 & 228 & 180 & 305 & 260\\
        \hline
        6 & 274 & 216 & 366 & 312 \\
        \hline
        7 & 320 & 252 & 427 & 364\\
        \hline
        \end{tabular}
    \caption{Comparison of resource count between the conditional adder circuit generated by \name and Qiskit.}
    \label{fig:gate_count}
    \end{minipage}
\end{figure}

\paragraph{Synthesis with \name}
We let \name synthesize a program with the specification below
\begin{align*}
\ket{f[1]}\ket{0}\ket{A[n]}\ket{B[n]}\mapsto \ket{f}\ket{f*c_0}\ket{A}\ket{f*A+B}
\end{align*}
The term $f*A+B$ indicates qubit $q_0$ is the flag qubit. With this specification, \name generates program $S$ shown in Fig~\ref{fig:cadder}. We compare the resource count between the conditional adder circuits generated by \name and Qiskit, which is shown in Fig~\ref{fig:gate_count}. All circuits are decomposed with $\{u_3, CNOT\}$ gateset by Qiskit's decomposition pass for comparison, where $u_3$ is a generic single-qubit rotation gate. The conditional adder programs generated by \name always use fewer quantum resources compared to the one from Qiskit.
\subsection{Eigenvalue Inversion}
\paragraph{Background and Motivation}
Eigenvalue inversion is a necessary arithmetic step in HHL\cite{lloyd2010quantum}, a quantum algorithm for linear systems of equations. Given a $c$-qubit eigenvalue state $\ket{\lambda}$, the eigenvalue inversion circuit needs to calculate $\ket{1/\lambda}$. In practice, only the first $n$ decimal places of the $1/\lambda$, denoted as $d_n$, and the remainder $r_0$ are kept. The precision $n$ depends on the accuracy requirement of the algorithm. 

\paragraph{Synthesis with \name}
We let \name synthesize a program that can calculate the first $n$ decimal places of $1/\lambda$ and keep the remainder $r_0$ for further use. 

Since \name's specification syntax only supports binary integers, we use the fact $2^n=\lambda*d_n+r_0$ where $d_n$ is the quotient we want to give the specification. The example below shows our intuition.
\begin{align*}
&\text{\textbf{Base 10:}} &1/7 = 0.\textcolor{red}{142857}14...  &\Leftrightarrow 10^6/7 = \textcolor{red}{142857}.14... &\Leftrightarrow & 10^6 = \textcolor{red}{142857}*7 + r_0\\
    &\text{\textbf{Base 2:}} &1/7 = 0.\textcolor{red}{001001}001...  &\Leftrightarrow  2^6/7 = \textcolor{red}{001001}.001... &\Leftrightarrow 
 & 2^6 = (\textcolor{red}{001001})_2*7 + r_0
\end{align*}
The specification we send to \name for the $n$-bit precision eigenvalue inversion program is
\begin{align*}
\ket{0}\ket{\Lambda[c]}\ket{0_{n}}\ket{1}\ket{0_{c-1}}\mapsto \ket{0}\ket{\Lambda}\ket{2^{n} \% \Lambda}_c\ket{2^{n} / \Lambda}
\end{align*}
The specification uses $\Lambda[c]$ to represent the signed input $c$-bit eigenvalue $\lambda$. It suggests qubit $q_{c+1}\sim q_{2c}$ store the remainder (i.e. $\ket{2^{n} \% \Lambda}$) and qubit $q_{2c+1} \sim q_{2c+n}$ store the quotient (i.e. $\ket{2^{n} / \Lambda}$). 

\begin{figure}
    \centering
    \begin{minipage}[m]{0.38\textwidth}
\begin{lstlisting}[language=python]
def inversion(N):
	circ = QuantumCircuit(N+2*c+1)
	def S(circ, n):
		if n<1:
			pass
		else:
			Append(circ,SUB(c),0,1,c+n+1)
			Append(circ,C_ADD(c-1),
        2*c+n,c,c+n)
			circ.x(2*c+n)
			circ=S(circ,n-1)
		return circ
	circ=S(circ,N)
	return circ
\end{lstlisting}
    
    \caption{$N$-bit precision eigenvalue inversion program.}
    \label{fig:inversion}
    \end{minipage}~
    \begin{minipage}[m]{0.57\textwidth}
    \includegraphics[width=\textwidth]{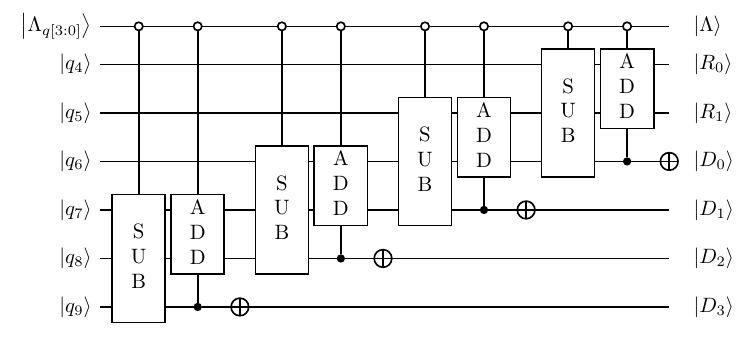}
    \caption{Eigenvalue inversion circuit for $n=3,c=3$.}
    \label{fig:inver_circ}
    \end{minipage}
\end{figure}

With this specification, \name generates the program shown in Fig~\ref{fig:inversion}. Fig~\ref{fig:inver_circ} shows the corresponding circuit for $n=3,c=3$. This program is a variant of Thapliyal~\cite{thapliyal2019quantum}'s general quantum division circuit. Compared to calculating $\ket{1/\lambda}$ with Thapliyal's division circuit, which is constructed by $max(n,c)$-qubit subtractor and conditional adder, this program uses $c$-qubit one. This significantly reduces the number of qubits required when $n$ is large.

\subsection{Quantum Fourier Transform}
\paragraph{Motivation and Background}
Quantum Fourier Transform (QFT)~\cite{coppersmith2002approximate} is the classical discrete Fourier Transform applied to the vector of amplitudes of a quantum state. QFT is a part of many quantum algorithms, notably Shor's algorithm~\cite{shor1994algorithms}, QPE algorithm~\cite{kitaev1995quantum}, and algorithms for the hidden subgroup problem~\cite{ettinger1999quantum}. A QFT over $\mathbb{Z}_{2^n}$ can be expressed as a map in two equivalent forms:
\begin{align}
    \text{QFT: } \ket{x} \mapsto \frac{1}{\sqrt{2^n}} \sum_{y=0}^{2^n-1} e^{\frac{2\pi i\cdot xy}{2^n}} \ket{y} \qquad \text{OR}\qquad     \text{QFT: } \ket{x} \mapsto \bigotimes_{k=0}^{n-1} \ket{z_k} \\
    \ket{z_k} = \frac{1}{\sqrt{2}}(\ket{0} + e^{2\pi i \cdot [0.x_kx_{k-1}...x_0]}\ket{1}),\qquad
    [0.x_m...x_1x_0] = \sum_{k=0}^m \frac{x_k}{2^{m-k+1}}.
    \label{equ:zn}    
\end{align}
\paragraph{Synthesis with \name}
We use the specification in Equation~\ref{equ:qftspec} to synthesize a program $S_{Q}$ that $S_Q(n)$ returns the $n+1$-qubit QFT circuit.

\name fails to synthesize a simple fixpoint structure QFT circuit. Therefore we synthesize it in two steps to help \name synthesize a nested structure circuit.
First, we let the synthesizer generate a program $S_z$ that transforms the state of qubit $n$ into state $\ket{z_n}$ and keep the state of qubits $q_0\sim q_{n-1}$ unchanged. The specification is:
\begin{align*}
    \ket{X[n+1]}\mapsto   \ket{X_0X_1...X_{n-1}} \otimes (\ket{0}\uplus e^{\frac{2\pi i \cdot X}{2^n}}\cdot\ket{1})
\end{align*}

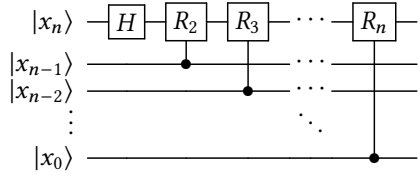
\begin{figure}
    \centering
    \begin{minipage}[m]{0.4\textwidth}
\begin{lstlisting}[language=Python]
def Zn(N):
    circ = QuantumCircuit(N+1)
    def S(circ,n):
        if n == 0:
            circ.h(N)
        else:
            S(circ,n-1)
            circ.cp(pi/2**n , N-n, N)
    S(circ,N)
    return circ
\end{lstlisting}
    \end{minipage}
    ~
    \begin{minipage}[m]{0.48\textwidth}
        
\centerline{
\Qcircuit @C=0.8em @R=0.75em {
   \lstick{\ket{x_{n}}}   &   \gate{H}  &   \gate{R_{2}}   &   \gate{R_{3}}  &   \qw & \cdots&        &   \gate{R_n}   &   \qw          \\
   \lstick{\ket{x_{n-1}}}   &   \qw       &   \ctrl{-1}     &   \qw           &   \qw &   \cdots&      &   \qw           &    \qw    \\
   \lstick{\ket{x_{n-2}}}   &   \qw       &   \qw           &   \ctrl{-2}     &   \qw &   \cdots  &   &  \qw            &   \qw         \\
   \lstick{\vdots }         &             &                 &                 &    &   \ddots &     &                 &                 \\
   &&&&&&&& \\
   \lstick{\ket{x_{0}}}     &   \qw       &   \qw           &   \qw           &   \qw &   \qw  &  \qw  &   \ctrl{-5}     &    \qw    
}}
    \end{minipage}
    \caption{Qiskit program \texttt{Zn} that transforms qubit $q_n$ to state $\ket{z_n}$}
    \label{fig:CRZN}
\end{figure}

With this specification, the synthesizer generates the program as shown in Fig~\ref{fig:CRZN}. Statement \texttt{circ.cp(pi/2**n, N-n, N)} in Qiskit means a controlled phase rotation gate $R_n$\footnote{Precisely, $R_n$ is a single-qubit unitary  
 $\begin{pmatrix}
1 & 0\\
0 & \omega_{n}
\end{pmatrix}$
where 
$\omega_n = exp(2\pi i / 2^n)$.

} on qubit $q_N$ controlled by qubit $q_{N-n}$.

Then we insert this \sqirs program $S_z$ (i.e. Qiskit program $\texttt{Zn})$ into the database so \name can use it for further synthesis, which leads to the QFT program in Fig~\ref{fig:QFT}.
%and Fig~\ref{fig:QFT-circ}.
\begin{figure}
    \centering
    \begin{minipage}[m]{0.42\textwidth}
\begin{lstlisting}[language=python]
def QFT(N):
    circuit = QuantumCircuit(N+1)
    def S(circ,n):
        if n == 0:
            circ.h(n)
        else:
            circ.append(Zn(n))
            S(circ, n-1)
    S(circuit, N)
    return circuit
\end{lstlisting}
    % \begin{align*}
    %     S_Q&:= \fix_1~ Id~ (\const{\texttt{H 0}})~(\const{CRZN})~(\const{I})
    % \end{align*}
    
    \caption{$N+1$-bit QFT program written in Qiskit language.}
    \label{fig:QFT}
    \end{minipage}~\hspace{2em}
    \begin{minipage}[m]{0.42\textwidth}
\begin{lstlisting}[language=python]
def teleport(N):
    circuit=QuantumCircuit(3*N)
    def S(circ, n):
        if(n<0):
            return
        else:
            circ.h(N+n)
            circ.cx(N+n, 2*N+n)          
            circ.cx(n, N+n)
            circ.h(n) 
            S(circ,n-1)
    S(circuit,N)
    return circuit
\end{lstlisting}
    % \begin{align*}
    %     S_Q&:= \fix_1~ Id~ (\const{\texttt{H 0}})~(\const{CRZN})~(\const{I})
    % \end{align*}
    
    \caption{$N$-qubit quantum teleportation program.}
    \label{fig:teleport}
    \end{minipage}
    % \begin{minipage}[m]{0.48\textwidth}
    % \small
    % \begin{tabular}{|c|c|}
    % \hline
    %      \textbf{Benchmark} & \textbf{Time (s)} \\
    %      \hline
    %     GHZ & 0.6\\
    %     Full Adder & 47.6\\
    %     Cuccaro Adder & 78.4\\
    %     Subtractor & 87.8\\
    %     Conditional Adder & 80.2\\
    %     Eigenvalue Inversion & 14.5\\
    %     QFT & 55.1 (step 1), 8.2 (step 2) \\
    %     \hline
    %     \end{tabular}
    % \caption{\name's runtime in seconds of the synthesis for example cases.}
    % \label{tab:perform}
    % \end{minipage}
\end{figure}
\subsection{$n$-qubit Quantum Teleportation}
\paragraph{Motivation and Background}

\begin{wrapfigure}{R}{.4\textwidth}
\includegraphics[width=0.4\textwidth]{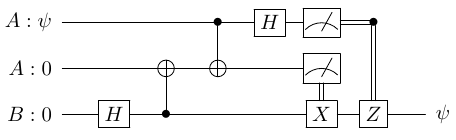}
\end{wrapfigure}
Quantum teleportation\cite{bennett1993teleporting} is one of the most famous quantum applications that can be implemented in the near term. It is a technique for transferring quantum information from a sender at one location to a receiver some distance away. The sender does not have to know the particular quantum state being transferred. Moreover, the recipient's location can be unknown, but to complete the quantum teleportation, classical information needs to be sent from sender to receiver. 
The right-hand side circuit shows the process for sending one-qubit state $\ket{\phi}$ from Alice to Bob.  
\paragraph{Synthesis with \name}
We let \name synthesize the unitary circuit part (before the measurement) of the $n$-qubit quantum teleportation process.
Suggest Alice wants to send a state $\ket{\phi_n}$ stored in $n$ data qubits to Bob, and they each have $n$ more ancilla qubits which are initialized to state $\ket{0}^n$ for the teleportation process. We follow the same measurement strategy as the one-qubit teleportation: Alice will measure $\ket{\phi_n}$ and Alice's $n$ aniclla qubits after the unitary circuit, then Bob apply bit-wise CZ and CX gate to Bob's $n$ ancilla qubits based on the result. Assume the data qubits, Alice's ancilla qubits and Bob's ancilla qubits are in state $\ket{a_n}, \ket{b_n}, \ket{c_n}$ respectively after the measurement, Bob can use bit-wise CZ and CX gate to reproduce the state $\ket{\phi_n}$ if $(-1)^{\oplus a_n} (\ket{b_n} \oplus \ket{c_n}) = \ket{\phi_n}$. Here $\ket{b_n} \oplus \ket{c_n}$ is bit-wise XOR operation (e.g. $\ket{101} \oplus (\ket{100} + \ket{110}) = \ket{001} + \ket{011}$) and $\oplus a_n$ is a reduction XOR operation on $a_n$ (e.g. $\oplus 101 = 1 \oplus 0 \oplus 1 = 0)$. With this intuition, we let \name use the specification below to synthesize the unitary circuit part of a $n$-qubit quantum teleportation program,
\begin{align*}
    \ket{\phi[n]}\ket{0_n}\ket{0_n}\mapsto \biguplus_{z \in \{0,1\}^{3n}} \delta(\phi = b_n\oplus c_n) \cdot e^{-\pi i \cdot (\oplus a_n)} \ket{z}
    \end{align*}
where $a_n = z[n-1:0], b_n = z[2n-1:n], c_n = z[3n-1, 2n]$. Fig~\ref{fig:teleport} shows the program generated by \name.
\subsection{Performance Evaluation}
\label{sec:performance}
In this section, we compare the performance of \name{} and previous circuit synthesis methods.

\paragraph{Implementation}
In the experiment, we use the Syntax-Guided Top-down Tree Search~\cite{sumit2017foundations} as the searcher with the following bounds on the search space: (1)
when searching a candidate program under \sqirs syntax in Definition~\ref{def:syntax}, 
we set the maximum program length to 10 and enumerate the value of $k$ in the \textsc{FIX} syntax in $\{1,2,3\}$; (2)
shorter candidate programs are sent to the verifier first; (3)
when searching a permutation $\pi$ under the syntax in Definition~\ref{def:permutation}, 
we set the maximum syntax derivation depth to 4; (4) when deriving the syntax rule $e::=m, m\in \nat$, we enumerate $m \in [0,3]$. 
All benchmarks in this paper can be synthesized under these bounds. 

The implementation of \name uses 1k lines of Python.
All the experiments are run with Z3 solver version 4.8.9 and Python 3.8.
\paragraph{Benchmarks}
Table~\ref{tab:benchmark} summarizes all \numbenchmark{} benchmarks. They are in three categories: arithmetic circuits,
state preparation and sub-programs widely used in quantum algorithms. Many benchmarks are collected from textbooks~\cite{nielsen_chuang_2010}.
Arithmetic circuits are frequently used in quantum oracle designs which is necessary for most famous quantum algorithms 
(e.g. Simons, Shor’s algorithms, Grover search algorithm). State preparation is necessary for the setup of many quantum
 applications (e.g. quantum teleportation). We also collect the necessary quantum sub-program used in quantum algorithms
 from their paper (e.g. HHL algorithm).

\begin{table}[h]
    \centering 
    \caption{Summary of all benchmarks used in the evaluation.}
    \label{tab:benchmark}
    \small
    \begin{tabular}{C{5em}C{5em}C{18em}C{8em}}
    \hline
    \textbf{Benchmark Type} & \textbf{Benchmark Name} & \textbf{Description} & \textbf{Gate Set} \\
    \hline
    \hline
      \multirow{3}{*}{\shortstack{State\\ Preparation}}  & n-GHZ & Greenberger–Horne–Zeilinger state~\cite{greenberger1989going}  & H,CX \\

      \cline{2-4}
      & n-Uniform & $n$-qubit uniform distribution state & H, X, Y, Z\\
      \hline
      \multirow{6}{*}{Arithmetic} & n-full-Add  & $n$-qubit full adder &QFT, CX, SWAP, CCX, X\\
      \cline{2-4}
      & n-Add & $n$-qubit in place adder & MAJ, UMA, QFT, CS, CZ, X\\
      \cline{2-4}
      & n-Sub & $n$-qubit in place subtractor & MAJ, UMA, QFT, CS, CZ, X\\
      \cline{2-4}
      & Cond n-Add &$n$-qubit conditional in place adder & Toffoli, QFT, CX, CS, CZ, X\\
       \hline
      \multirow{5}{*}{\shortstack{Algorithm\\ Module}}& n-QFT & $n$-qubit Quantum Fourier Transform  & H,CS,CT, SW AP \\
      \cline{2-4}
      & Inversion & $n$-qubit precision eigenvalue inversion for HHL algorithm~\cite{lloyd2010quantum} & c-adder, subtractor,X\\
      \cline{2-4}
      & n-Toff & $n$-qubit Toffoli gate  & Toffoli, CX, X \\
    \cline{2-4}      
      &n-Teleport & $n$-qubit quantum teleportation & H, CX\\

      \hline
    \end{tabular}
\end{table}

We compare the performance of \name against QFAST~\cite{younis2021qfast} and Qsyn~\cite{kang2023modular}.
For \name, we use input-output style specification written in \name-spec as input.
For the other two frameworks, we use their specification interfaces and try to synthesize circuits with $n=3,4,5,6$. We use the same gate set when comparing \name and Qsyn in each case, while for QFast we use its hard-coded gate set.

We stop the synthesis and regard the synthesis as a failure if the running time is over 1 hour. 
All runtimes are a median of three runs.  

\begin{table}[h]
    \centering 
    \caption{Running time of all benchmarks used in the evaluation.}
\label{tab:performance}
    \small
    \begin{threeparttable}
    \begin{tabular}{c|c|cccc|cccc}
    \hline
     \textbf{Benchmark} & \textbf{\name} & \multicolumn{4}{c|}{\textbf{QFAST time (s)}} & \multicolumn{4}{c}{\textbf{Qsyn time (s)}} \\
     \textbf{Name}& \textbf{time (s)} & 3 & 4 & 5 & 6 & 3 & 4 & 5 & 6 \\
    \hline
    \hline
       n-GHZ & 1.793 & 0.122 & 2.48 & 59.1& -\tnote{*} & 0.091 & 1.33 & 49.2 & -\tnote{*}  \\
       n-Uniform & 0.415 & 0.027 & 1.25 & 32.7& -\tnote{*} & 0.096 & 1.40 & 39.8 & -\tnote{*}\\
      \hline
        n-Full-Add & 288.1 & 617  & -\tnote{*}& -\tnote{*} & -\tnote{*} & 812.4 &-\tnote{*} & -\tnote{*} & -\tnote{*}\\
n-Add & 170.4  & 942 & 3511 & -\tnote{*} & -\tnote{*} &  132& 2818 & -\tnote{*} & -\tnote{*}\\
       n-Sub & 168.6 & 667 & -\tnote{*} & -\tnote{*} & -\tnote{*} & 287 & 3240 & -\tnote{*} & -\tnote{*}  \\
       Cond n-Add & 185.2 & 851 & -\tnote{*}& -\tnote{*} & -\tnote{*} & 192 & 1804 & -\tnote{*} & -\tnote{*}\\
      \hline
       n-Toff & 66.7  &<0.01 & 21.8 & 679 & -\tnote{*} & <0.01 & 16.5 & 354 & -\tnote{*} \\
      n-QFT& 145.49 & 21.4 & 397 & 3598 & -\tnote{*}  & 67.2 & 473 & -\tnote{*} & -\tnote{*}  \\
       Inversion & 93.5 & 1622 & -\tnote{*} & -\tnote{*} & -\tnote{*} & 22.5 & 740 & -\tnote{*} & -\tnote{*} \\
       n-Teleport & 185.1 & 897 & -\tnote{*} & -\tnote{*} & -\tnote{*} & 614 & -\tnote{*} & -\tnote{*} & -\tnote{*} \\

      \hline
    \end{tabular}
    \begin{tablenotes}
\footnotesize
      \item{*} Time out after 1 hour.
\end{tablenotes}
\end{threeparttable}
\end{table}

Table~\ref{tab:performance} shows the running time of all the experiments. 
We can see that \name{} successfully synthesizes programs in all \numbenchmark{} benchmarks in at most 5 minutes, while QFAST and Qsyn fail to synthesize circuits for $n = 6$. From the result, we can see that when the size increases, the time of QFAST and Qsyn indeed grow exponentially, while \name{} only pays a fixed cost for all sizes. 

The synthesis time of \name{} is comparable to the time to synthesize a corresponding circuit with size 3, with an exception of the n-Toff benchmark. We note that this is because 3-Toffoli is exactly a single Toffoli gate and 4-Toffoli can be done using two Toffoli gates, which are straightforward for QFAST and Qsyn to search. In contrast, \name{}'s search is longer because it needs to consider the inductive structure and corner cases of $n=1$ and $2$. Nevertheless, \name{} quickly outperforms other frameworks on n-Toff at $n=5$.

\section{RELATED WORK} 
\label{sec:relatedwork}
\paragraph{Synthesis of quantum circuits}
Many methods have been proposed to synthesis quantum circuits of a fixed size~\cite{shende2006synthesis, de2020methods, amy2013meet, saeedi2011synthesis, kitaev1997quantum,deng2023automating, younis2021qfast,kang2023modular,xu2023synthesizing}. These methods do not consider the inductive structure of quantum programs and do not scale due to their exponential blowup with the number of qubits.

\paragraph{Synthesis of classical programs}
The tasks of synthesizing classical programs are intensively-studied in the recent decades~\cite{sumit2017foundations, kitzelmann2009inductive}. The problem definitions of program synthesis are diverse and orienting, including syntax-guided synthesis \cite{jha2010oracle, alur2013syntax, alur2018search, hu2018syntax}, example-guided synthesis \cite{gulwani2011automating, gulwani2012spreadsheet, polozov2015flashmeta}, semantics-guided synthesis \cite{kim2021semantics, d2021programmable}, and resource-guided synthesis \cite{knoth2019resource, hu2021synthesis}. The modern approaches to solve these problems make use of sophisticated search algorithms, such as enumerative search with pruning \cite{gulwani2011synthesizing, phothilimthana2016scaling}, constraint solver like satisfiability modulo theory (SMT) solver \cite{jha2010oracle, feng2017component, solar2008program}, and machine learning \cite{liang2010learning, menon2013machine}. 
We refer curious readers to surveys~\cite{sumit2017foundations, kitzelmann2009inductive} for a comprehensive picture on the development of classical program synthesis techniques. Many synthesis frameworks are developed into productive tools. For example, the \textsc{Sketch} \cite{solar2008program} framework completes programs with holes by specifications. \textsc{Rosette} \cite{torlak2013growing} builds solvers into the language to automatically fill in holes when programming. However, \name needs to deal with unique challenges from quantum programs. 

\paragraph{Verification of quantum programs}
An important procedure in syntax-guided synthesis is to verify any candidate program. Various logic and verification tools for quantum programs are developed in the last decade. QWIRE~\cite{paykin2017qwire} embedded the formal verification of quantum programs manually in the Coq proof assistant. QBricks \cite{qbricks2020deductive} do formal verification of quantum programs semi-manually using Why3. Quantum abstract interpretation~\cite{yu2021quantum} provides efficient tools to test the properties of quantum programs. In particular, the path-sum representation \cite{amy2018towards} of quantum program semantics inspired our representation. %We employ SMT solvers to check the correctness of searched programs. 
\citet{chen2023automata} uses tree automaton to verify fixed-size quantum circuits. It is hard to generalize to general-size cases because graphical structures like automaton are hard to symbolically model in SMT solvers.

Quantum Hoare logic ~\cite{ying2012floyd} uses quantum predicates and Hoare triples to express and derive properties of quantum programs. Its language, quantum while language does not have the detailed structure of unitary executions. Its specifications are quantum predicate matrices. Therefore, it cannot be applied to synthesize quantum unitary circuit families.

The use of SMT solvers to automate the reasoning has also appeared in Giallar~\cite{tao2022giallar}, Quartz~\cite{xu2022quartz} and symQV~\cite{symQV}, although they only work on quantum circuit compilation passes, quantum circuit optimizations or fixed-size quantum circuit verification.

%In the regime of quantum computing, circuit synthesis \cite{shende2006synthesis, de2020methods, amy2013meet, saeedi2011synthesis, kitaev1997quantum, younis2021qfast} means constructing quantum circuits according to a unitary.
%As instances, the Solovay-Kitaev theorem \cite{kitaev1997quantum} gives an algorithm approaching any unitary with a universal set of quantum gates. QFAST \cite{younis2021qfast} synthesizes a target unitary by enumerative searches with pruning. Though dealing with similar synthesis tasks, we focus on synthesizing \emph{a family of circuits} represented by a succinct quantum program satisfying specifications. 

\section{DISCUSSION AND FUTURE WORKS}
\label{sec:future}

\name comes with several limitations. At a high level, \name sacrifices the expressiveness of the specification due to the limitation in efficient SMT encodings. 
For example, the lack of the equivalence verification of general complex functions forces us to consider the special form of amplitude in Definition~\ref{def:ppsa}.
The sparsity requirement of $\alpha$ is another such restriction. 
Any relief of such restrictions would enlarge the space of programs that can be synthesized by \name.

Another limitation is that \name cannot directly synthesize programs involving nested fixed-point structures. Synthesizing a program nested loop or fixed-point structure is also a challenging problem in the classical domain. 
This is because the loop invariant of the program in nested loop structures is unknown and usually non-trivial to figure out. 
So we make the current \name only expand the \textsc{FIX} syntax at most once when generating the candidate program and avoid directly sending a program in nested fixed-point structure to the \name verifier. 
A natural next step is to include the search for the loop invariant in nested loop structures as part of \name, and to verify the candidate program against both the specification and the generated loop invariant.

Many quantum algorithms such as Bernstein-Vazirani~\cite{bv} and Deustch-Jozsa~\cite{dj} have quantum oracle as part of the program. However, it is hard to synthesize and verify quantum programs with oracles because it requires higher-order logic to quantify over arbitrary oracles. It is also an interesting next step to extend \name{} to support quantum oracles.

The design of \sqirs{} inherits concrete qubit indices from \sqir{}. 
This design introduces complications in the inductive variant that have to be addressed with explicit permutations and relabelings. We will explore the possibility of using a different representation of variables to make synthesis less complicated in the future.

Besides extending the expressiveness of specification and the support of more complicated quantum programs, it is also interesting to 
improve the synthesis performance by integrating classical program synthesis techniques into the quantum domain, including counterexample-guided synthesis,
and various search heuristics.

\section{CONCLUSION}
\label{sec:conclusion}
We present \name, the first quantum program synthesis framework,  including a new inductive quantum programming language, its specification, a sound logic for reasoning, and an encoding of the reasoning procedure into SMT instances. 
By leveraging existing SMT solvers, \name successfully synthesizes seven quantum unitary programs and \name can generate programs better than the standard solutions in the quantum subtractor and quantum conditional adder cases. These programs can be readily transpiled
to executable programs on major quantum platforms, e.g., Q\#, IBM Qiskit.

\name constitutes the first step toward a fully automated quantum program synthesis framework, which can significantly ease the task of programming in dealing with the low-level details, and hence leave the human programmers to focus on the high-level design of the system. 

\begin{acks}                           
We thank anonymous reviewers for constructive suggestions that improve the presentation of the paper. H.D., Y.P., and X.W. was partially funded by the U.S. Department of Energy, Office of Science, Office of Advanced Scientific Computing Research, Quantum Testbed Pathfinder Program under Award
Number DE-SC0019040, Air Force Office of Scientific Research under award number FA9550-21-1-0209, the U.S. National Science Foundation grant CCF-1942837 (CAREER) and a Sloan research fellowship.
\end{acks}

\section*{Data-Availability Statement}
QSynth is available from Zenodo DOI 10.5281/zenodo.10054966~\cite{QSynthZen} and the latest version from \href{https://github.com/sqrta/QSynth}{Github}

\bibliographystyle{ACM-Reference-Format}
\bibliography{references}
\newpage
% Appendix
\appendix
\section{Appendix}\
\label{sec:appendix}
\subsection{Proofs}
\label{app:proofs}

Proof of \theom{soundness}

\begin{proof}
We verify each rule's soundness by reasoning about the semantics of the program constructs.

\paragraph{\textsc{Weaken}} For $(n, x, y)\in h'\subset h,$ there is $\bra{y}\sem{\ssem{S}(n)}\ket{x}=\alpha(n, x, y)$, hence $\interpret{h}{S}{\alpha}.$

\paragraph{\textsc{Const}} For any $(n, x, y)\in\mathbb{N}^3,$ note $\ssem{\const{P}}(n)=P$ and $\alpha(n, x, y)=\sem{P}(x, y).$ This makes $\bra{y}\sem{\ssem{\const{P}}(n)}\ket{x}=\alpha(n,x,y)$, and $\interpret{\mathbb{N}^3}{\const{P}}{\alpha}.$

\paragraph{\textsc{Replace}} For any $(n, x, y)\in h,$ since $\bra{y}\sem{\ssem{S}(n)}\ket{x}=\alpha(n, x, y)=\alpha'(n, x, y),$ we conclude $\interpret{h}{S}{\alpha'}.$

\paragraph{\textsc{Relabel}} For any $(n, x, y)\in h,$ there is $\bra{y}\sem{\ssem{S}(n)}\ket{x}=\alpha(n, x, y).$ Because $\pi(n)$ is injective, for any $(n, u, v)\in\pi\circ h$, there exists $x, y$ such that $\pi(n, x)=u$ and $\pi(n, y)=v$. Then $\bra{u}\sem{\ssem{\relabel{\pi}{S}}(n)}\ket{v}=\bra{\pi(n, y)}\sem{\ssem{\relabel{\pi}{S}}(n)}\ket{\pi(n, x)}=\bra{y}\sem{\ssem{S}(n)}\ket{x}=\alpha{n, x, y}=\alpha'(n, u, v).$

\paragraph{\textsc{Seq}} For any $(n, x, y)\in h,$ note $\bra{y}\sem{\ssem{\seq{S_1}{S_2}}(n)}\ket{x}=\bra{y}\sem{\ssem{S_2}(n)}\sem{\ssem{S_1}(n)}\ket{x}$
$=\sum_z\bra{y}$ $\sem{\ssem{S_2}(n)}\ket{z}\bra{z}\sem{\ssem{S_2}(n)}\ket{x}.$ For any $z$, if $(n, x, z)\in h_1 \wedge \alpha_1(n, x, z)=0$, then $\bra{z}\sem{\ssem{S_2}(n)}\ket{x}=\alpha_1(n, x, z)=\alpha_1(n, x, z)\alpha_2(n, z, y)$. Similarly, if $(n, z, y)\in h_2 \wedge \alpha_2(n, z, y)=0$, it equals to $\alpha_1(n, x, z)\alpha_2(n, z, y)$. If $(n, x, z)\in h_1\wedge (n, z, y)\in h_2,$ the term also becomes $\alpha_1(n, x, z)\alpha_2(n, z, y).$ Hence we have $\interpret{h}{\seq{S_1}{S_2}}{\alpha}.$

\paragraph{\textsc{Fix}} Similarly, we denote $\fix_k~\pi~P_0~\cdots~P_{k-1}~S_L~S_R$ as $\fix_k$. For any $(i, x, y)\in h$ where $i<k$, by $(h_i, \alpha_i)\equiv_i (h, \alpha)$, we know $(i, x, y)\in h_i$ and $\alpha(i, x, y)=\alpha_i(i, x, y)$. Since $\interpret{h_i}{\const{P}_i}{\alpha_i}$ and $(h, \alpha)\equiv_i(h_i, \alpha_i),$ we have $\bra{y}\sem{\ssem{\fix_k}(i)}\ket{x}=\bra{y}\sem{P_i}\ket{x}=\alpha_i(i, x, y)=\alpha(i, x, y).$ For any $(i, x, y)\in h$ such that $i\geq k$, note $\ssem{\fix_k}=\ssem{S_L}(i); \texttt{map\_qb}(\pi(i),\ssem{\textbf{fix}_k}(i-1));\ssem{S_R}(i).$ According to the proof for \textsc{Relabel} and \textsc{Seq}, with $(h, \alpha)\equiv_i (h_L, \alpha_L)\otimes(\pi\circ\text{pred }(h, \alpha))\otimes(h_R, \alpha_R),$ we have $\bra{y}\sem{\ssem{\fix_k}(i)}\ket{x}=\alpha(i, x, y).$
\end{proof}

~\\

\noindent Proof of Theorem~\ref{theo:sqir}
\begin{proof}
We first prove that given a \sqir program $P$ and we have $\interpret{\nat^3}{\const{P}}{\alpha}$, then $\alpha$ is sparse.

Since $P$ is a \sqir program, $P$ is a sequence of applications of gates to fixed number of qubits and we have
\begin{align}
    \alpha(n,x,y) = \ssem{P}_{xy}
\end{align}
Without loss of generality, suppose $\ssem{P}$ is a unitary applied to qubits $q_0,q_1,...,q_m, m\in \nat$. Let function $\X,\Y$ be
\begin{align}
    \X(n,y)= \begin{cases}
            \{k \mid k<2^{m+1}, k\in \nat\},~ y<2^{m+1}\\
            \{\mask(y,0,m) + k \mid k<2^{m+1}, k\in \nat\},~ \text{otherwise}
    \end{cases}\\
    \Y(n,x)= \begin{cases}
            \{k \mid
            k<2^{m+1}, k\in \nat\},~ x<2^{m+1}\\
            \{\mask(x,0,m) + k \mid k<2^{m+1}, k\in \nat\},~ \text{otherwise}
    \end{cases}
\end{align}
where $\mask(x,0,m)$ means to set from the  0th bit to the $m$th bits (in the order from low to high) in $x$ to 0 (e.g. $\mask((1111)_2,0,1))=(1100)_2=(12)_{10})$. It is easy to see that for arbitrary inputs, the size of the sets returned by $\Y,\X$ is always $2^{m+1}$.
Now we prove that
\begin{align}
   \forall n,x,y \in \nat,\alpha(n,x,y)\neq 0 \rightarrow y\in \Y(n,x)  
\end{align}
Formula $\alpha(n,x,y)\neq 0 \rightarrow x\in \X(n,y)$ can be proved in the same way. When $\alpha(n,x,y)\neq 0$:
\begin{itemize}
    \item if $x<2^{m+1}$, since $\ssem{P}$ is a unitary applied to qubits $q_0,q_1,...,q_m$, $\bra{x}\ssem{P}\ket{y}=0$ for every $y\ge 2^{m+1}$. So we know that $y<2 ^{m+1}$ and $y\in \Y(n,x)$.
    \item Otherwise, since $\ssem{P}$ does not change the state of qubits other than qubits $q_0,q_1,q_2,...,q_m$, we should have $\mask(x,0,m) = \mask(y,0,m)$ if $\alpha(n,x,y)\neq 0$. Then we have $y\in \Y(n,x)$
\end{itemize}
So we have $\sparse{\alpha}{\X}{\Y}$ and we prove that $\alpha$ is sparse.\par

Next we prove that given a sparse amplitude function $\alpha$ and an injective mapping $\pi$, function $\pi\circ\alpha$ is sparse.

Since $\alpha$ is sparse, there exists functions $\X, \Y$ and we have $\sparse{\alpha}{\X}{\Y}$. Let functions $\X^\pi, \Y^\pi$ be
\begin{align}
    \X^\pi(n,y)=\{ \pi^{-1}(k) \mid k\in \X(n,\pi(y))\}\\
    \Y^\pi(n,x)=\{ \pi^{-1}(k) \mid
         k\in \Y(n,\pi(x))\}
\end{align}
Then we have
\begin{align*}
    (\pi\circ \alpha)(n,x,y)\neq 0 &\Longrightarrow\alpha(n,\pi(x),\pi(y))\neq0 \\ &\Longrightarrow \pi(x)\in \X(n,\pi(y))\land \pi(y)\in \Y(n,\pi(x)) \\ 
    &\Longrightarrow x\in \{ \pi^{-1}(k) \mid k\in \X(n,\pi(y))\} \land y\in \{ \pi^{-1}(k) \mid k\in \Y(n,\pi(x)\}\\ &\Longrightarrow x\in \X^\pi(n,y) \land y\in \Y^\pi(n,x) 
\end{align*}
So we have
\begin{align*}
    \forall n~x~y, (\pi\circ \alpha)(n,x,y)\neq 0 \rightarrow y\in \Y^\pi(n,x) \land x\in \X^\pi(n,y)
\end{align*}
So we have $\sparse{,\pi \circ \alpha}{\X^\pi}{\Y^\pi}$ and we prove that $\pi \circ \alpha$ is sparse.

Finally we prove that given two sparse amplitude function $\alpha_1,\alpha_2$, function $\alpha_1 * \alpha_2$ is sparse. Since $\alpha_1,\alpha_2$ are sparse, suppose we have $\sparse{\alpha_1}{\X_1}{\Y_1}$ and $\sparse{\alpha_2}{\X_2}{\Y_2}$. We also have
\begin{align*}
        (\alpha_1*\alpha_2)(n,x,y) = \sum_{z \in \X_1(n,y)} \alpha_1(n,x,z) \alpha_2(n,z,y)
\end{align*}
Let functions $\X, \Y$ be
\begin{align}
    \X(n,y) := \{\X_1(n,z) | z\in \X_2(n,y)\}\\
    \Y(n,x) := \{\Y_2(n,z) | z \in \Y_1(n,x) \} 
\end{align}
Then we have
\begin{align}
    (\alpha_1*\alpha_2)(n,x,y)\neq 0 &\Longrightarrow \exists z, \alpha_1(n,x,z)\neq 0 \land \alpha_2(n,z,y)\neq0\\
    & \Longrightarrow \exists z,  z\in \Y_1(n,x) \land z \in \X_2(n,y) \land x \in \X_1(n,z) \land y \in \Y_2(n,z)\\
    & \Longrightarrow x \in \X(n,y) \land y\in\Y(n,x)   
\end{align}

So we have
\begin{align}
    \forall n~x~y, (\alpha_1*\alpha_2)(n,x,y)\neq 0 \rightarrow x \in \X(n,y) \land y\in\Y(n,x)
\end{align}
So $\sparse{\alpha_1*\alpha_2}{\X}{\Y}$ and we prove that $\alpha_1*\alpha_2$ is sparse.
\end{proof}

~\\

\end{document}